\documentclass[aip,reprint]{revtex4-1}

\usepackage{graphicx}
\usepackage{dcolumn}
\usepackage{bm}

\usepackage[utf8]{inputenc}
\usepackage[T1]{fontenc}
\usepackage{mathptmx}
\usepackage{etoolbox}
\usepackage{amsmath}
\usepackage{amssymb}
\usepackage{gensymb}
\usepackage{graphicx}
\usepackage{soul}
\usepackage{upgreek}
\usepackage{esint}
\usepackage[normalem]{ulem}
\usepackage{times}
\usepackage{setspace}
\usepackage{url}
\usepackage{bigints}
\usepackage{xcolor}
\usepackage{array}

\makeatletter
\def\@email#1#2{%
 \endgroup
 \patchcmd{\titleblock@produce}
  {\frontmatter@RRAPformat}
  {\frontmatter@RRAPformat{\produce@RRAP{*#1\href{mailto:#2}{#2}}}\frontmatter@RRAPformat}
  {}{}
}%
\makeatother

\begin{document}

\title[Balancing Hodge Laplacians]{Balanced Hodge Laplacians Optimize Consensus Dynamics over Simplicial Complexes}

\author{Cameron Ziegler}
\affiliation{Department of Mathematics, University at Buffalo, State University of New York, Buffalo, NY 14260}
\author{Per Sebastian Skardal}%
\affiliation{ 
Department of Mathematics, Trinity College, Hartford, CT 16106
}

\author{Haimonti Dutta}
\affiliation{Management Science and Systems Department, University at Buffalo, State University of New York, Buffalo, NY 14260
}

\author{Dane Taylor}
\email{danet@buffalo.edu}
\affiliation{Department of Mathematics, University at Buffalo, State University of New York, Buffalo, NY 14260}

\date{\today}

\begin{abstract}
Despite the vast literature on network dynamics, we still lack basic insights into dynamics on higher-order structures (e.g., edges, triangles, and more generally, $k$-dimensional ``simplices'') and how they are influenced through higher-order interactions.  A prime example lies in neuroscience where groups of neurons (not individual ones) may provide the building blocks for neurocomputation. Here, we study consensus dynamics on edges in simplicial complexes using a type of  Laplacian matrix called a Hodge Laplacian,  which we generalize to allow higher- and lower-order interactions to have different strengths. Using techniques from algebraic topology, we study how collective dynamics converge to a low-dimensional subspace that corresponds to the homology space of the simplicial complex. We use the Hodge decomposition to show that higher- and lower-order interactions can be optimally balanced to maximally accelerate convergence, and that this optimum coincides with a balancing of dynamics on the curl and gradient subspaces. We additionally explore the effects of network topology, finding that consensus over edges is accelerated when 2-simplices are well dispersed, as opposed to  clustered together.
\end{abstract}

\maketitle

\section{Introduction}\label{sec:intro} 

Interest in dynamical processes occurring on and through higher-order structures in simplicial complexes has grown rapidly in recent years, with examples including random walks \cite{schaub2020random, mukherjee2016random, rosenthal2014simplicial}, social contagions \cite{iacopini2019simplicial, matamalas2020abrupt}, neuronal activity \cite{petri2014homological}, cellular networks \cite{vergne2014simplicial}, synchronization \cite{gambuzza2020master, skardal2019abrupt, arnaudon2021connecting}, and consensus \cite{maletic2014consensus, neuhauser2020multibody}. Such dynamics are useful for applications ranging from biological processes \cite{arai2014effects} to the design of machine learning \cite{roddenberry2019hodgenet,bunch2020simplicial} and signal processing  \cite{schaub2021signal}. 
In this pursuit, a variety of mathematical tools have been developed ranging from mean field theory \cite{gleeson2012accuracy} and probability theory \cite{snijders2010introduction}, to persistent homology \cite{bardin2019topological}, algebraic topology \cite{hatcher2005algebraic}, and  Hodge theory \cite{muhammad2006control}. Thus, new problems in modeling and understanding the dynamical systems on simplicial complexes are revitalizing interest in analytical methods from algebraic topology and Hodge theory.

Consensus over networks is a popular model for  collective decision  making  in   cognitive \cite{hinsz1990cognitive}, social \cite{fiol1994consensus}, and biological systems \cite{conradt2005consensus}, and more recently, it has been used to investigate  interconnected human-AI decision systems \cite{song2021asymmetric}. 
Notably, consensus dynamics provide a foundation for decentralized optimization algorithms,
which strategically implement consensus using synchronous or asynchronous gossiping between nodes  \cite{tsitsiklis1986distributed,boyd2005gossip}.
Such algorithms are often employed to take advantage of distributed computing infrastructure to more efficiently train machine learning models, such as support vector machines \cite{bijral2017data,huynh2021} and deep neural networks \cite{assran2019stochastic,niwa2020edge,vogels2020powergossip,kong2021consensus}. 
For such systems, each node trains a local model on  local data, and at the same time,   communication between nodes  enables them to reach a consensus on what the model parameter should be. 
Notably, much of the literature on decentralized learning 
aims to understand and optimize the converge rate to consensus,
which subsequently minimizes the time  and number of computations that are required to train a predictive model.

Herein, we extend the study of consensus to simplicial complexes by assigning states to higher-dimensional simplices and allowing interactions through both their lower and higher-dimensional faces. 
Our work complements other recent extensions of consensus to simplicial complexes \cite{bianconi2021topological, millan2020explosive, gambuzza2020master} in which states assigned to each node interact through higher-dimensional simplicies. 
Contrasting this prior work, we propose a model that utilizes a generalization of the Hodge Laplacian matrix in which the relative influence of higher and lower-order interactions can be tuned via a \emph{balancing parameter} $\delta$.
This leads to our definition of a \emph{generalized Hodge Laplacian}, and our work is motivated by a recent similar matrix called a normalized Hodge Laplacian \cite{schaub2020random}. 
    
With the help of an associated subspace decomposition called the \emph{Hodge Decomposition}, we provide a convergence analysis of our model and use tools from algebraic topology to investigate the role of \emph{simplicial complex homology}. We find that the collective dynamics converge to a low-dimensional subspace lying within the homology subspace, which includes and generalizes the notion of connected components by way of Betti numbers (i.e., connected components characterize 0-dimensional homology). Furthermore, we show that higher- and lower-order interactions can be optimally balanced to maximally accelerate convergence, and that this optimum coincides with a balancing of dynamics on the curl and gradient subspaces of a Hodge decomposition, respectively. We additionally explore the effects of topology, finding that consensus over edges is accelerated when 2-simplices are well dispersed, i.e., as opposed to  clustered together.  

Our findings provide new insights into how  dynamical systems with higher-order interactions can be engineered and optimized, which not only improves our understanding of  self-organizing  biological and physical systems, but may also support the development of improved consensus-based algorithms for machine learning. Specifically, 
recent work \cite{roddenberry2019hodgenet,bunch2020simplicial} has illustrated advantages for neural networks that use simplicial complexes and Hodge Laplacians. Our formulation, analysis, and optimization of consensus over simplicial complexes may help  
future research implement decentralized versions of such emerging frameworks by allowing them to also take advantage of decentralized optimization techniques that rely on  consensus.

This paper is organized as follows. 
We present background information in Sec.~\ref{sec:background},  our model and the role of homology  in Sec.~\ref{sec:results}, optimization theory in \ref{sec:optimization}, and a discussion in Sec.~\ref{sec:discussion}.

\section{Background information}\label{sec:background}

We provide here background information on the homology of simplicial complexs (Sec.~\ref{sec:back_simplicial}) as well as Hodge Laplacians and decompositions (Sec.~\ref{sec:back_hodge}).

\subsection{Simplicial Complexes, Boundary Matrices and Homology}\label{sec:back_simplicial}

We begin with some definitions.
\emph{Simplicial complexes}   generalize undirected graphs (which encode dyadic interactions) by allowing polyadic interactions among sets of nodes/vertices.  They consist of \textit{simplices} of any number of dimensions. For a set $\mathcal{V}$ of nodes, a $k$-dimensional simplex (or simply, $k$-simplex) $\mathcal{S}\subset \mathcal{V}$ is a subset with $k+1$ elements.  We  will refer to $0$-simplices as \textit{nodes} or \textit{vertices},   $1$-simplices  as  \textit{edges}, and 2-simplices as ``filled in'' \textit{triangles}. 
For a $k$-simplex $\mathcal{S}$, any $(k-1)$-simplex that is a subset of $\mathcal{S}$ with $k$ elements is called its  \textit{face}. For any such face, $\mathcal{S}$ is called its   \textit{coface}. For example, an edge $(i,j)$ has two faces, nodes $i$ and $j$, and $(i,j)$ is a coface to both  $i$ and $j$. 
Two $k$-simplices are called \textit{lower adjacent} if they share a face, and they are called \textit{upper adjacent} if they share a coface.  For example, two edges are lower adjacent if they share a common node, and the edges are upper adjacent if they are two sides of the ``boundary'' of 2-simplex. (We will formally define boundaries below.)

A \textit{simplicial complex} is a set of simplices in which any face of a  simplex must also be contained in the simplicial complex, and the intersection between faces is either another simplex or an empty set. The dimension of a simplicial complex is the maximum dimension of its simplices. For example, a 1-dimensional simplicial complex contains  only nodes and edges and is equivalent to an undirected graph.  For simplicity, herein we will focus on 2-dimensional simplicial complexes; however, we emphasize that our results naturally extend to higher dimensions. 

Many of our calculations require simplicial complexes with \textit{oriented simplices}. An orientation of a $k$-simplex is an equivalence class of the ordering of its nodes, where two classes are equivalent if they are the same up to an even number of permuations. The most convenient way to orient a simplicial complex is to order its vertices $x_1, x_2, \dots x_n$.  From this ordering, for all $k > 1$ choose the orientation of each $k$-simplex $\{x_{i_0}, x_{i_1}, \dots x_{i_{k}}\}$ to be $[x_{i_0}, x_{i_1}, \dots x_{i_{k}}]$ such that $i_0 < i_1 < \dots < i_{k}$. Note that this ordering is done only for bookkeeping purposes, and does not represent a direction to the simplices.

Importantly, connectivity of faces and cofaces can be encoded using boundary maps and their associated boundary matrices.
Their precise definition stems from studying functions that are defined over over $k$-simplices, and we call such a function a \textit{$k$-chain}, i.e., a linear combination of $k$-simplices. 
The set of all $k$-chains for a simplicial complex is the finite-dimensional vector space $C_k$. We can then define \textit{boundary maps} $\partial_k: C_k \rightarrow C_{k-1}$ by their action on a simplicial complex: 
\begin{equation}
    \label{boundary operator}
    \partial_k([x_0, x_1, \dots, x_{k}]) = \sum_{i=0} (-1)^i [x_0, \dots, x_{i-1}, x_{i+1}, x_k]
\end{equation}
This takes a simplex to an alternating sum of its faces.  Representing the map $\partial_k$ gives us the \textit{boundary matrix} $B_k$, with columns corresponding to the $k$-simplices and rows corresponding to the $(k-1)$-simplices.  An entry $[B_k]_{ij}$ of $B_k$ is $\pm1$ if the $i$th $(k-1)$-simplex is  a face of the $j$th $k$-simplex (where the sign indicates orientation), and it is otherwise 0. Boundary matrices fully encode a simplicial complex in that it can be recreated from its boundary matrices.

\begin{figure}[h!]
  \includegraphics[width=.7\linewidth]{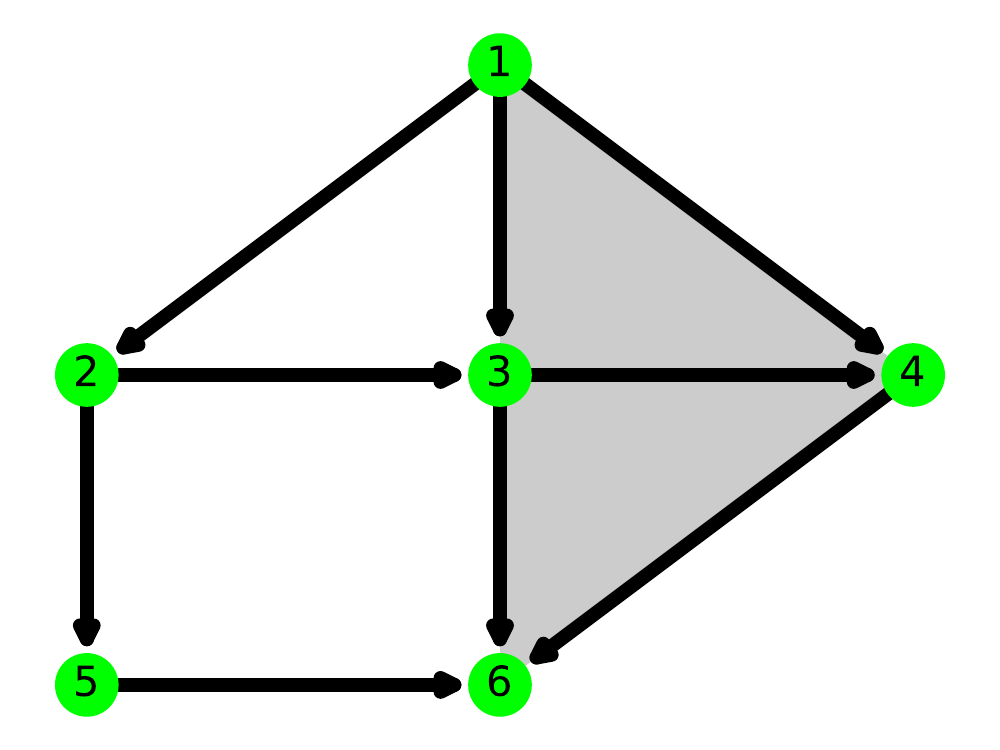}
  \\ \vspace{.4cm}
  {\small 
  $B_1 = \begin{tabular}{ c|c c c c c c c c c c }
  & [1, 2] & [1, 3] & [1, 4] & [2, 3] &[2, 5] & [3, 4] & [3, 6] & [4, 6] & [5, 6] 
  \\ \hline
  1 & -1 & -1 & -1 & 0 & 0 & 0 & 0 & 0 & 0 \\
  2 & 1 & 0 & 0 & -1 & -1 & 0 & 0 & 0 & 0 \\
  3 & 0 & 1 & 0 & 1 & 0 & -1 & -1 & 0 & 0 \\
  4 & 0 & 0 & 1 & 0 & 0 & 1 & 0 & -1 & 0 \\
  5 & 0 & 0 & 0 & 0 & 1 & 0 & 0 & 0 & -1 \\
  6 & 0 & 0 & 0 & 0 & 0 & 0 & 1 & 1 & 1 \\
\end{tabular}$
}
\\
\vspace{.4cm}
$B_2 = \begin{tabular}{ c|c c} 
  & [1, 3, 4] & [3, 4, 6]
\\ \hline
  \lbrack1, 2\rbrack & 0 & 0 \\
  \lbrack1, 3\rbrack & 1 & 0 \\
  \lbrack1, 4\rbrack & -1 & 0 \\
  \lbrack2, 3\rbrack & 0 & 0 \\
  \lbrack2, 5\rbrack & 0 & 0 \\
  \lbrack3, 4\rbrack & 1 & 1 \\
  \lbrack3, 6\rbrack & 0 & -1 \\
  \lbrack4, 6\rbrack & 0 & 1 \\
  \lbrack5, 6\rbrack  & 0 & 0 \\
\end{tabular}$
  \caption{
  {\bf Example simplicial complex and its boundary matrices}.
  A simplicial complex with 6 nodes, 9 edges, and 2 triangles. The nodes are indexed from 1 to 6. The edges of the simplicial complex have arrows not because they are directed edges, but to show their chosen orientation, where the arrow points to the node with a larger index. Its boundary matrices are given by $B_1$ and $B_2$.}
  \label{fig:boundary}
\end{figure}

In Fig.~\ref{fig:boundary}, we illustrate an example simplicial complex and its boundary matrices.  Note that the edges are undirected, and arrows only indicate the simplices' chosen orientations. Also note that it is a 2D simplicial complex, since  there are no simplices of dimension $3$ or greater (i.e., all higher dimensional boundary matrices are empty).
Observe that each column of $B_1$ corresponds to an edge $(i,j)$,  and all entries in a column are zero except for the $i$-th and $j$-th entries, which contain $-1$ and $1$, respectively. The rows and columns of $B_2$  correspond to edges and 2-simplices, respectively, and values $\pm 1$ indicate incidences between faces and cofaces. The signs indicate whether or not the edges' orientations match that for a triangular cycle/boundary around each 2-simplex.  For example, the orientation of edge $(1, 3)$ matches that for 2-simplex $(1, 3, 4)$. In contrast, the orientation of $(1, 4)$ does not since the cycle would be completed by traversing from node 4 to node 1 and not 1 to 4.

Of note, this simplicial complex has  four separate cycles:   two cycles are  the boundary of a 2-simplex,  $\{(1,3),(3,4),(4,1)\}$ and $\{(3,4),(4,6),(6,3)\}$;
and two cycles are not such a boundary:  $\{(1,2),(2,3),(3,1)\}$ and $\{(2,3),(3,6),(6,5),(5,2)\}$. 

The study of cycles (and higher-dimensional ``holes'' including voids)  generalizes the study of connected components in graphs and is the focus of homology.
With these boundary functions and a simplicial complex $X$, we can define the \textit{$n$th (simplicial) homology group} $H_n(X)$ as Ker$(\partial_n)/$Im$(\partial_{n+1})$.  The dimension of the $H_n$ is called the \textit{$n$th Betti number}.  These Betti numbers provide us with important topological information about simplicial complexes. The $0$th Betti number is equal to the number of connected components in the simplicial complex, while the $1$st gives us the number of cycles that are not ``filled-in'' by triangles.  Higher-dimensional Betti numbers  indicate the number of higher-dimensional ``holes'' in a simplicial complex \cite{hatcher2005algebraic}.

\begin{figure*}[t]
  \includegraphics[width=.95\linewidth]{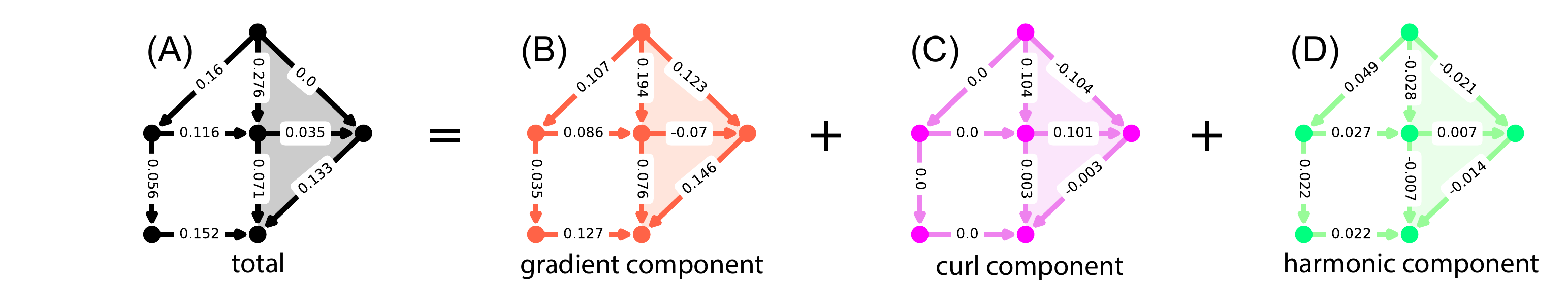} 
  \caption{
  {\bf Hodge decomposition of a real-valued function defined over 1-simplices.}  
  (A) A 1-chain ${\bf x}\in\mathbb{R}^{n^{k}}$ that is defined over $n^{k}=9$   edges (i.e., 1-simplices).
  (B) Projection ${\bf x}^{(g)}$ onto the gradient subspace.
  (C) Projection ${\bf x}^{(c)}$ onto the curl subspace.
  (D) Projection ${\bf x}^{(h)}$ onto the harmonic subspace.  
  }
  \label{fig:hodge_decomp}
\end{figure*}

\begin{figure*}[t]
  \includegraphics[width=.95\linewidth]{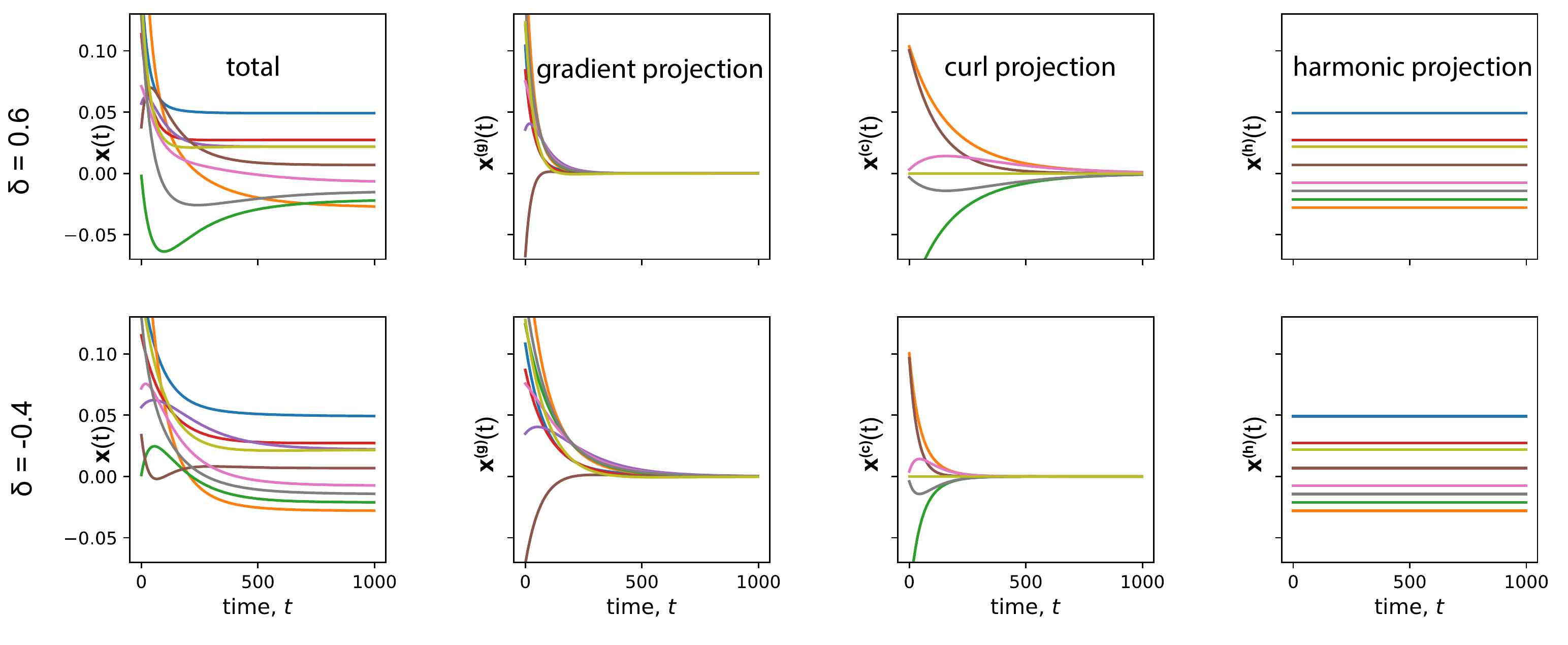} 
  \caption{
  {\bf Balancing parameter $\delta$ influences convergence rate of GHL-1 consensus dynamics}.
  Two simulations of GHL-1 consensus were run on the simplicial complex from Fig. \ref{fig:boundary}.  The top row shows the Hodge decomposition of the dynamics when $\delta = 0.6$, while the bottom row has $\delta = -0.4$.  Both simulations began with the same initial condition.  Notice that the curl subspace converges faster when $\delta$ is negative, while the gradient subspace converges faster when $\delta$ is positive.  As expected, the harmonic subspace does not change under these dynamics, and so the total function converges to the harmonic projection.}
  \label{fig:timeseries}
\end{figure*}

\subsection{Hodge Laplacians and Hodge Decomposition}\label{sec:back_hodge}

The $k$-th combinatorial Hodge Laplacian   
\begin{equation}
    \label{combinatorial_Hodge}
    L_k = B_k^T B_k + B_{k+1} B_{k+1}^T
\end{equation}
is a symmetric, positive semi-definite, and square matrix of size  $n_k$, where $n_k$ is the number of $k$-simplices in the simplicial complex. From this definition, we have that $B_k^T B_k$ is calculated from the lower-adjacencies of $k$-simplices, while $B_{k+1} B_{k+1}^T$ is from their upper-adjacencies. Finally, notice that $L_0$ is recovers the widely known  combinatorial Laplacian of a graph: $L_0 = B_1 B_1^T = D-A$, where $A$ is the adjacency matrix and $D$ is a diagonal matrix that encodes node degrees, $D_{ii}=\sum_jA_{ij}$.   

If for some $k$, we consider maps that give each $k$-simplex some real value (i.e. real-valued $k$-chains), the space of all such maps is $\mathbb{R}^{n_k}$, where $n_k$ is the number of $k$-simplicies in the simplicial complex. Using the combinatorial Hodge Laplacian $L_k$, we can decompose that space $\mathbb{R}^{n_k}$ into three mutually orthogonal subspaces \cite{lim2015hodge}.
\begin{equation}
    \label{Hodge decomposition}
    \mathbb{R}^{n^k} =  \text{im}(B_k^T)  \oplus \text{im}(B_{k+1}) \oplus \text{ker}(L_k)
\end{equation}  
This is called the \textit{Hodge decomposition}, and these three subspaces are known as the \textit{gradient, curl}, and \textit{harmonic} subspaces, respectively. From this, we can immediately see that the gradient subspace only depends on the $(k-1)$-dimensional simplicies, the curl subspace depends on the $(k+1)$-dimensional simplicies, and the harmonic subspace depends on both. 
Importantly, the harmonic space and homology group are isomorphic vector spaces \cite{lim2020hodge}, and  the Betti number is the dimension of both spaces.

In Fig.~\ref{fig:hodge_decomp}(A), we set $k=1$ and consider a random real-valued function ${\bf x}$ that is defined on the edges of the simplex that was shown Fig.~\ref{fig:boundary}. In panels (B), (C) and (D) we depict, respectively, the projection  of ${\bf x}$ onto the gradient, curl and harmonic subspaces. We denote these ${\bf x}^{(g)}$, ${\bf x}^{(c)}$ and ${\bf x}^{(h)}$, respectively. Examining Fig.~\ref{fig:hodge_decomp},  we highlight several important properties of these subspaces. 
First, the gradient components sum to zero along any cycles. 
Second, the curl components are zero for edges that are not a 2-simplex boundary, and the entries sum to zero around each node.  
Third, the harmonic component sums to zero around each node, and it also sums to zero along each 2-simplex (but not the other cycles) \cite{schaub2020random}.

\section{Consensus over Simplicial Complexes with weighted Hodge Laplacians  }\label{sec:results}

\subsection{Model for Consensus}\label{sec:model}

Herein, we present a new formulation for consensus dynamics over simplicial complexes using Hodge Laplacians. However, rather than utilize combinatorial Hodge Laplacians, we instead define and utilize \textit{Generalized Hodge Laplacians} (GHLs): 
\begin{equation}
    \label{weighted_Hodge}
    L_k^{(\delta)} = (1+ \delta) B_k^T B_k + (1-\delta) B_{k+1} B_{k+1}^T .
\end{equation}
Specifically, we introduce a \textit{balancing parameter} $\delta \in [-1, 1]$ to account for the possibility that for some applications, it is reasonable to assume that the higher- and lower-dimensional interactions should have different strengths of influence. Notably, interactions through lower-adjacencies (e.g., via shared nodes when $k=1$) are more strongly weighted when $\delta>0$, whereas interactions through higher-adjacencies (e.g., via shared 2-simplices when $k=1$) are more strongly weighted when $\delta<0$.
For the choice $\delta=0$, the GHL $L_k^{(\delta)}$ recovers the combinatorial Hodge Laplacian, i.e., $L_k^{(0)}=L_k$.

Importantly, we point out the relevance of the Hodge decomposition is exactly the same for GHLs as for the combinatorial Hodge Laplacian. That is because our introduction of $\delta$ in Eq.~\eqref{weighted_Hodge} is just a scalar re-weighting of the two summands in the right-hand side of Eq.~\eqref{combinatorial_Hodge}, and these matrices' image spaces are orthogonal and unchanged by $\delta$.
We point out that this ``preservation'' of the Hodge decomposition contrasts another recent generalization called \emph{normalized Hodge Laplacians} \cite{schaub2020random}. That is,   normalization  uses the degrees (i.e., number of neighbors) of $k$-simplicies of different dimensions, and this can significantly alter (i.e., rotate) the Hodge subspaces.

Given a GHL, we consider a model for consensus over a simplicial complex using the following system of linear ordinary differential equations.
We define the linear dynamics 
\begin{equation}
    \label{dynamics}
    \frac{d}{dt}\textbf{x}(t) = -L_k^{(\delta)} \textbf{x}(t)
\end{equation}
as Generalized Hodge Laplacian-k (GHL-k) consensus. Note that $L_0^{(0)}=2L=2(D-A)$ again recovers the  combinatorial graph Laplacian, and so GHL-0 consensus with  $\delta=-1$ is $\frac{d}{dt}\textbf{x}(t) = -2L  \textbf{x}(t)$, proportional to the standard definition for consensus over a graph, $\frac{d}{dt}\textbf{x}(t) = -L  \textbf{x}(t)$ \cite{olfati2004consensus}.
In principle, one can define and study consensus over $k$-simplices of any dimension. However, in this work we will 
reduce our scope to the case when $k=1$ so that each edge $(i_p,j_p)$ is assigned a state $x_p\in\mathbb{R}$, and their states collectively evolve over time. For the remainder of this paper, we will identify  fundamental insights about dynamics evolving under Eq.~\eqref{dynamics} and the role that is played by the topology and homology of the simplicial complex as well as the associated Hodge subspace.

\subsection{Decoupling of Dynamics within Hodge Subspaces}\label{sec:decoupling}

We first show that the Hodge subspaces are invariant subspaces for the dynamics evolving under Eq.~\eqref{dynamics}.
Consider a solution   $\textbf{x}(t) = \textbf{x}^{(g)}(t) + \textbf{x}^{(c)}(t) + \textbf{x}^{(h)}(t)$ to Eq.~\eqref{dynamics}, where 
$\textbf{x}^{(g)}(t)$, $\textbf{x}^{(c)}(t)$, and $\textbf{x}^{(h)}(t)$, respectively, give the gradient, curl and harmonic components (i.e., projections onto the respective subspaces).
We substitute this form into Eq.~\eqref{dynamics} to obtain
\begin{widetext}
\begin{align}
    \label{decomposed dynamics}
    \frac{d}{dt}\left[ \textbf{x}^{(g)}(t) + 
    \textbf{x}^{(c)}(t) + 
    \textbf{x}^{(h)}(t) \right]
    &= -[L_k^{(\delta)}\textbf{x}^{(g)}(t) ]
    -[L_k^{(\delta)}\textbf{x}^{(c)}(t) ]
    -[L_k^{(\delta)}\textbf{x}^{(h)}(t)] \nonumber\\
    &= -[(1+\delta) B_1^T B_1 \textbf{x}^{(g)}(t)]
    -[(1-\delta)B_2 B_2^T \textbf{x}^{(c)}(t)]
    - [0] .
\end{align}
\end{widetext}

These simplifications use
\begin{itemize}
    \item $B_2 B_2^T \textbf{x}^{(g)}(t)=0$ since $\textbf{x}^{(g)}(t) \in \text{ker}(B_2B_2^T)$,
    \item $B_1^T B_1 \textbf{x}^{(c)}(t)=0$ since $\textbf{x}^{(c)}(t) \in \text{ker}(B_1^T B_1)$, \item and $L_1^{(\delta)} \textbf{x}^{(h)}(t)=0$ since $\textbf{x}^{(h)}(t) \in \text{ker}(L_1^{(\delta)})$.
\end{itemize}
Due to the orthogonality of these subspaces, we obtain the following  decoupled set of linear ODEs:
\begin{align}\label{dynamics decomposition2}
\frac{d}{dt}\textbf{x}^{(g)}(t)  &= -(1+ \delta) B_1^T B_1 \textbf{x}^{(g)}(t), \nonumber\\
\frac{d}{dt}\textbf{x}^{(c)}(t)  &= -(1-\delta)B_2 B_2^T \textbf{x}^{(c)}(t) ,\nonumber\\
\frac{d}{dt}\textbf{x}^{(h)}(t)&= 0,
\end{align}
which have the solutions
\begin{align}\label{decoupled solution}
    \textbf{x}^{(g)}(t) &= e^{-t(1+ \delta) B_1^T B_1} \textbf{x}^{(g)}(0), \nonumber \\
    \textbf{x}^{(c)}(t) &= e^{-t(1-\delta) B_2 B_2^T} \textbf{x}^{(c)}(0), \nonumber \\
    \textbf{x}^{(h)}(t) &= \textbf{x}^{(h)}(0)  .
\end{align}
In other words, the dynamics evolve separately within each Hodge subspace. Because the matrices $B_1^T B_1$ and $B_2 B_2^T$ are both positive semi-definite, the gradient and curl components converge to zero with increasing time, i.e., $\lim_{t\to\infty} {\bf x}^{(g)}(t) =  0$ and $\lim_{t\to\infty}  {\bf x}^{(c)}(t)=  0$. On the other hand, the harmonic component remains fixed in time, and so the full state converges to $\lim_{t\to\infty} {\bf x} (t) = {\bf x}^{(h)}(0)$, which we will call the \emph{consensus vector}.  

Of particular interest is the \emph{asymptotic convergence rate} $\mu(\delta)$ for Eq.~\eqref{dynamics}, which is the convergence rate for the slower subspace dynamics. That is, each subspace in the Hodge decomposition also converges, but not all at the same rate.   Let $\lambda_2^{(1)}$ and $\lambda_2^{(2)}$, be the smallest nonzero eigenvalues of $B_1^T B_1$ and $B_2 B_2^T$, respectively. The gradient subspace is dependent on the lower-dimensional interactions and  has convergence rate $(1+\delta) \lambda_2^{(1)}$.  The convergence curl subspace is similarly determined by the higher-dimensional interactions, and so its convergence rate is   $(1-\delta)\lambda_2^{(2)}$. Notably, the balancing parameter $\delta$ tunes the relative convergence rates for the  invariant dynamics in the gradient and curl subspace. It also follows that $\mu(\delta) = \min\{(1+\delta) \lambda_2^{(1)},(1-\delta)\lambda_2^{(2)}\}$.

In Fig.~\ref{fig:timeseries}, we see illustrate timeseries ${\bf x}(t)$ for simulations of Eq.~\eqref{dynamics} as well as the projections onto the three Hodge subspaces. The top and bottom rows indicate the same initial condition ${\bf x}(0)$ but two different choices for the balancing parameter: (top) $\delta=0.6$; and (bottom) $\delta=-0.4$.
Each curve indicates a trajectory $x_p(t)$ for some edge $(i_p,j_p)$ in the simplicial complex that was shown in the previous figures.
Observe in both rows that the gradient and curl components converge to zero, whereas the harmonic components remain fixed in time. In fact, for both values of $\delta$ the system states converge to the same consensus vector ${\bf x}^{(h)}(0)$---that is because the initial conditions are the same. That is, $\delta $ has no effect on the final state.
In contrast, $\delta$ has a significant impact on the convergence rates for the invariant dynamics within the gradient and curl subspaces.
Observe for significantly positive $\delta$ (i.e., when the lower-dimensional interactions are prioritized) that the gradient component converges much more quickly than the curl component.
The reverse is true when $\delta$ is sufficiently negative.

\begin{figure*}[t!]
    \includegraphics[width=\linewidth]{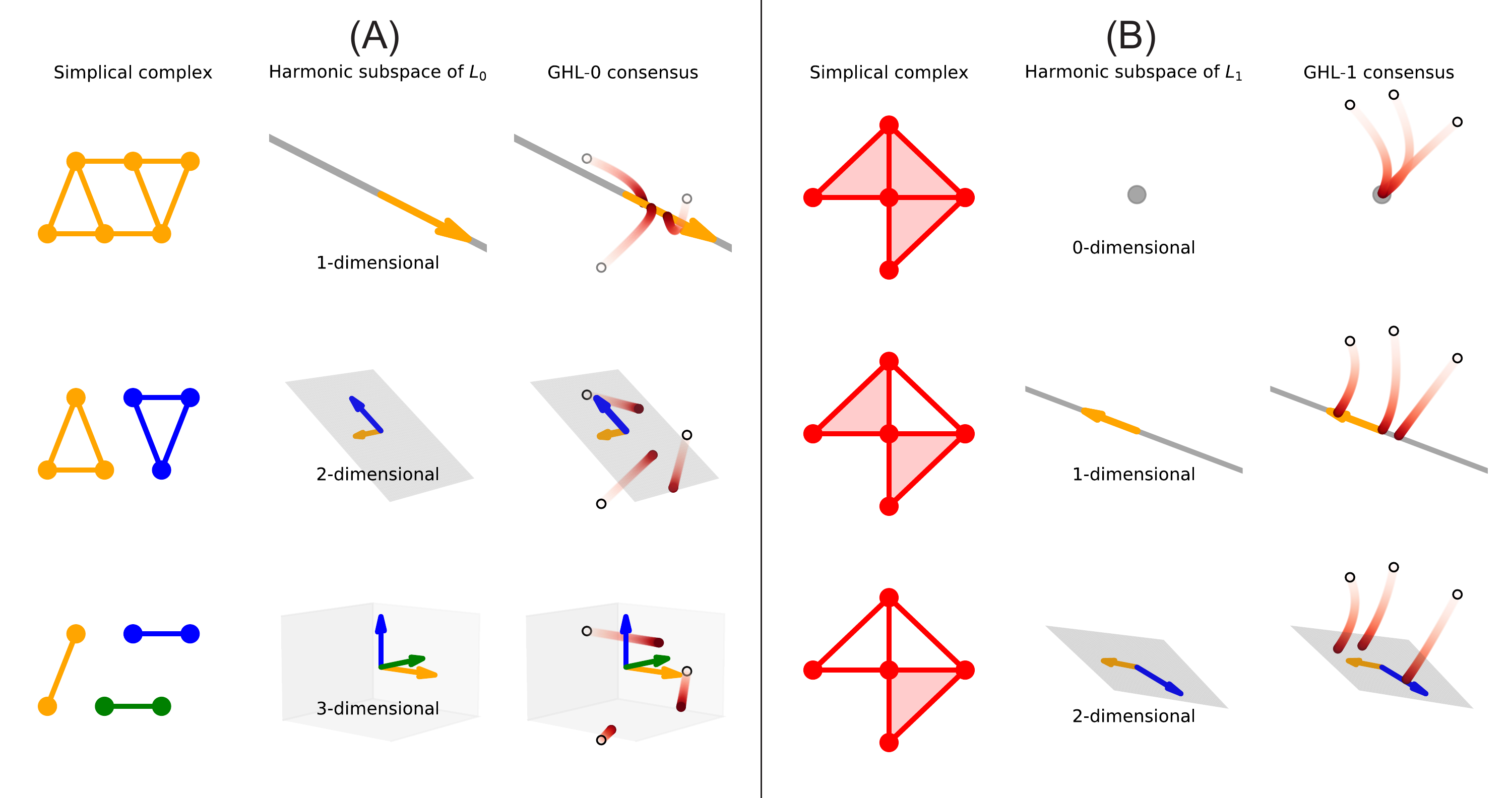}
  \caption{
  {\bf Consensus dynamics converge to the harmonic subspace that is isomorphic to the homology space}.
  The zero and first dimensional homology groups for a few example simplicial complexes.  The homology groups are illustrated as vector spaces.  We plot the convergence of GHL-0 (A) and GHL-1 (B) consensus with 3 different random initial conditions starting from the white circles.  Notice that they all converge to the harmonic subspace. Within each panel (A) and (B), the initial conditions are the same.  While the harmonic subspaces are plotted in full, the converging flows are projected down into 3 dimensions to simplify this visualization.}
  \label{fig:homology}
\end{figure*}

\subsection{Uncovering the Role of Homology}\label{sec:homology}

In the previous subsection, we saw that the consensus dynamics remains unchanged in the harmonic subspace.  That is to say, the consensus vector is   the harmonic component of the initial condition.  It's therefore important to take a look at properties of the harmonic subspace and their dependence on  the topology and homology of the simplicial complex.  
Importantly, the harmonic subspace and homology group are isomorphic vector spaces, and so the space of possible solutions to Eq.~\eqref{dynamics} is determined by the homology of the simplicial complex. In particular, the dimension, i.e., Betti number, gives the dimension of the solution space.

In Fig \ref{fig:homology}, we show several example simplicial complexes, their associated harmonic subspaces of $L_k$ (which are identical to those of $L_k^{(\delta)}$), and the convergence of GHL-k consensus to the harmonic space with either (A) $k=0$ or (B) $k=1$.

Starting with Fig \ref{fig:homology}(A), in the first column we depict  1-dimensional simplicial complexes such that Betti-0 (i.e., the dimension of both the harmonic and homology spaces  for $L_0$)  increases as one considers lower rows (see second column). Note that Betti-0 is   simply the number of connected components. In this case, the harmonic subspace of $L_0$ is spanned by vectors that have a constant value for entries associated with nodes in a given component and  all other entries are zeros. If there are $\beta_0$ such components, then there are $\beta_0$ such vectors that span $\text{Ker}(L_0)$. The third column illustrates that GHL-0 consensus---which is equivalent to the traditional definition of consensus over an undirected graph---converges to a consensus value in this harmonic subspace. For practical purposes, one typically assumes that the graph is strongly connected so that the harmonic space is a 1-dimensional line and the consensus value gives the mean value for the initial condition $\{x_p(0)\}$.

Turning our attention to  Fig \ref{fig:homology}(B), in the first column we depict 2-dimensional simplicial complexes such that Betti-1 (i.e., the dimension of both the harmonic and homology spaces for $L_1$)  increases as one considers lower rows (see second column). Observe that Betti-1  corresponds to the number of cycles that are not the boundary of a 2-simplex, i.e., ``unfilled triangles'' in these three examples.   The third column illustrates that GHL-1 consensus converges to a consensus value in the harmonic subspace $\text{Ker}(L_1)$. For some applications, we predict that it is beneficial for ${\bf x}(t)\to 0$. This   can be achieved by either choosing the initial condition to be orthogonal to the harmonic subspace,  ${\bf x}^{(h)}(0)=0$, or  ensuring that the simplicial complex has   a zero-dimensional harmonic subspace, i.e., no homology groups (i.e., unfilled cycles) in dimension $k$. 
For other applications---in particular, multiscale dynamical systems, it may also be beneficial for the dynamics to instead converge to a low-dimensional subspace, within which other dynamics could occur (e.g., oscillations and synchronization). Thus, although we call our proposed model GHL-k ``consensus'', we expect the findings that we present herein to also be relevant to much more diverse types of dynamics (see e.g., \cite{millan2020explosive,schaub2020random}).

\section{Optimizing Consensus Convergence Rate}\label{sec:optimization}

\begin{figure}[b]
\includegraphics[width=\linewidth]{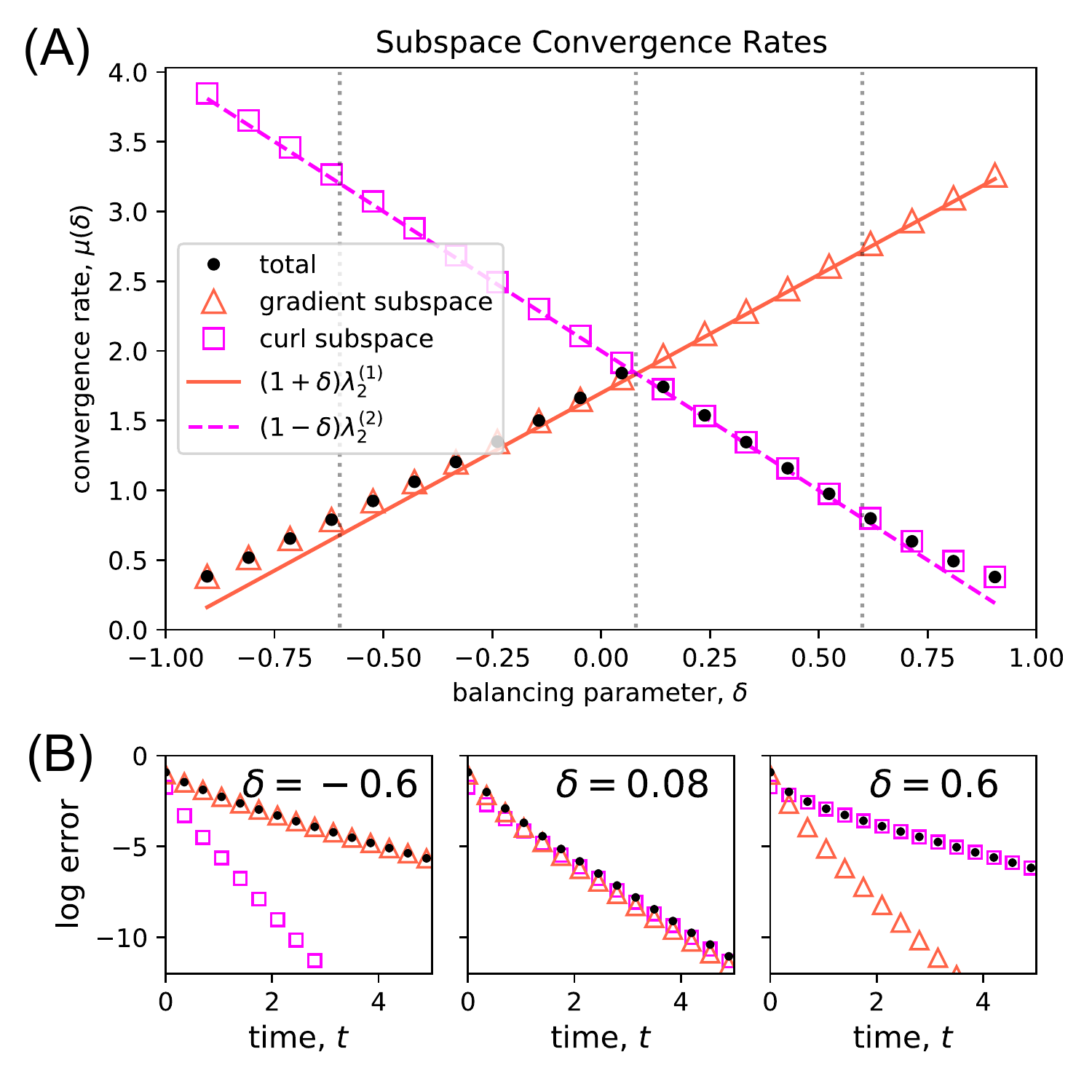}
\vspace{-.5cm}
\caption{
{\bf Balancing invariant subspace dynamics maximally accelerates convergence}.
Working with the simplicial complex from Fig. \ref{fig:boundary}, plot (A) shows the convergence rate of the total value as well as the gradient and curl subspaces in GHL-1 consensus simulations for different values of $\delta$. The orange and purple lines show the expected values for these convergence rates calculated using the eigenvalues of the component matrices of the generalized Hodge Laplacian as explained in Section \ref{sec:decoupling}.  For this simplicial complex, the balancing parameter that optimizes convergence rate is $\delta^* \approx .08$.  In (B), we plot the log normed error, $\log{(||{\bf x}(t)-\bf{x}^{(h)}||_2)}$, again for the total as well as the gradient and curl subspaces.  Remember that the consensus here converges to the harmonic projection of the initial condition.}
\label{fig:convergence}
\end{figure}

\subsection{Convergence Rate is Maximized by Optimal Balancing of Subspace Dynamics}

Recall the convergence rate of Eq.~\eqref{dynamics} is given by $\mu(\delta) = \min\{(1+\delta) \lambda_2^{(1)},(1-\delta)\lambda_2^{(2)}\}$ and that convergence occurs faster when $\mu$ is larger. We therefore seek to maximize the convergence rate by solving the following optimization problem

\begin{equation}
    \label{optimal delta} 
    \delta^* = \text{argmax}_{\delta} ~\mu(\delta) = \frac{\lambda_2^2- \lambda_2^1}{\lambda_2^2 + \lambda_2^1}.
\end{equation}

In this case, the optimum coincides with the intersection of two lines and is found by equating $(1+\delta) \lambda_2^{(1)} = (1-\delta)\lambda_2^{(2)}$. 
In other words,
the fastest total convergence requires that we choose the balancing parameter $\delta$ so that the decoupled dynamics with the gradient and curl subspaces have the same convergence rate. 
Such an optimum may be beneficial to engineering applications and the optimization of naturally occurring systems.

In Fig. \ref{fig:convergence}, we provide numerical validation for   such an optimum. 
The solid orange and dashed pink lines lines Fig.~\ref{fig:convergence}(A) indicate our theoretical predictions of convergence rate within the gradient and curl subspaces, respectively,. These are in excellent agreement with  empirically observed convergence rates for these two spaces, which are indicated by the orange triangles and pink squares, respectively. Black dots indicate empirically observed convergence rates for the full system, and note that they align the minimum of the two subspace convergence rates. 

Vertical dotted lines in Fig.~\ref{fig:convergence}(A) highlight three values of $\delta$ that are   chosen to be  below, at, and above the optimum $\delta^*\approx 0.08$. In Fig.~\ref{fig:convergence}(B), we plot empirically observed values of the error $|| {\bf x}(t) - {\bf x}(\infty)||$, in log scale, for the total system as well as   the gradient and curl components. Considering the left and right subpanels of Fig.~\ref{fig:convergence}(B), observe  that the gradient component converges more slowly than the curl component when $\delta < \delta^*$, and the opposite occurs when $\delta > \delta^*$. In either case, the slower dynamics slows the full system's convergence. As shown in the center panel of Fig.~\ref{fig:convergence}(B), the system converges fastest at the optimum $\delta^*$, where  the curl and gradient components have the same convergence rate.  Under this optimal choice for the  balancing parameter, we   say that $L_1^{(\delta^*)}$ is \emph{balanced}.

\begin{figure*}[t]
  \includegraphics[width=\linewidth]{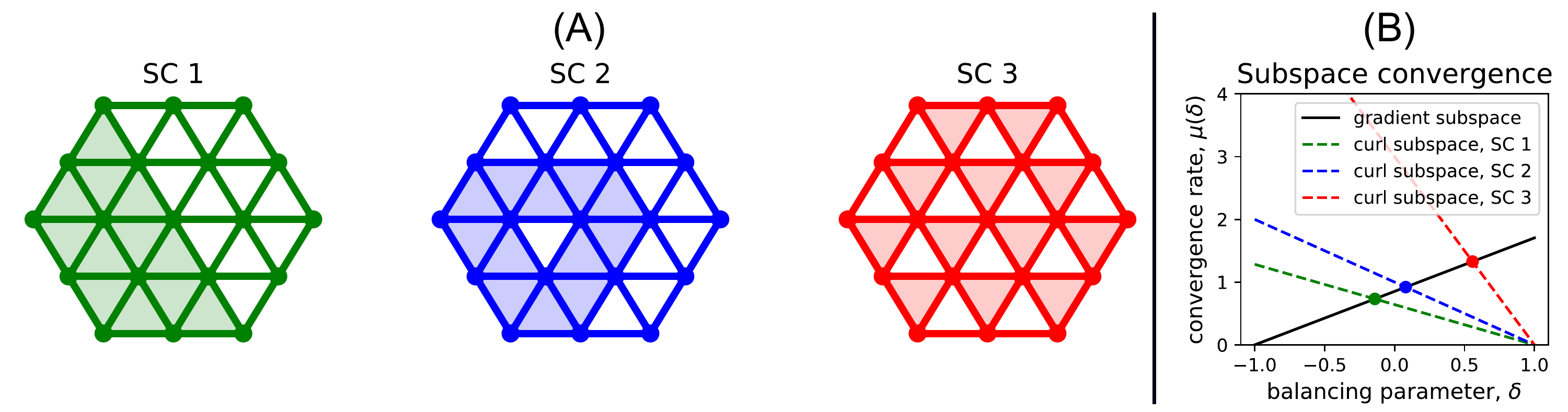} 
  \caption{\textbf{2-simplex arrangement affects convergence rate of curl component.} 
  (A) Simplicial complexes with different 2-simplices. Since they have identical nodes and edges, they have the same gradient subspaces but different curl subspace.
  (B) Convergence rate $\mu(\delta)$ versus $\delta$ for GHL-1 consensus on these three simplicial complexes.
  The three dots highlight how the optimum $\delta^*$ differs across the three simplicial complexes. Notably, convergence is fastest for SC 3 (i.e., where the 2-simplices are most dispersed), and the optimum occurs at a larger value of $\delta$.
  }
  \label{fig:exampleSC}
\end{figure*}
\subsection{Dispersed Higher-order Interactions Accelerate Consensus}

It is natural to wonder how the optimum $\delta^*$ is influenced by the topology and homology of a simplicial complex.  To this end, we created three example simplicial complexes that have the same underlying $1$-skeleton (the same nodes and edges), but different 2-simplices.  Specifically, each simplicial complex has the same number of triangles, but they have different placements, as shown in  Fig.~\ref{fig:exampleSC}(A).  Because they share the same nodes and edges, all three simplicial complexes have the same $B_1$ boundary matrix and associated gradient subspace. In contrast, their $B_2$ boundary matrices differ as well as their curl subspaces. 
In Fig \ref{fig:exampleSC}(B), we plot the convergence rates of the gradient and curl components for the three simplical complexes. First, observe that their gradient subspaces' convergence rates are all the same (see black line). In contrast, their curl subspaces have different convergence rates (see dashed lines). The three dots highlight the optimum $\delta^*$ for each simplicial complex. Observe that these dots move up and to the right as one considers simplicial complex 1, 2 and 3. That is, convergence occurs fastest  for simplicial complex 3 and slowest for simplicial complex 1. 
It follows that convergence is faster for the curl subspace (and the total system) when the 2-simplices are spaced apart, i.e., rather than clustered together.

Another interesting property is that while the optimal convergence rate $\mu(\delta^*)$ increases with each subsequent simplicial complexes (i.e., 1 to 3), the associated optimum $\delta^*$ also increases. Recall for our definition for the weighted Hodge Laplacian $L^{(\delta)}_k$ given by Eq.~\eqref{weighted_Hodge} that larger $\delta$ implies stronger coupling via lower-dimensional interactions (i.e., adjacency via shared nodes). 
In other words, because the total coupling weight is conserved [i.e., the weights are $(1+ \delta)$ and $(1-\delta)$], and because the curl subspace (which arises due to higher-dimensional interactions) converges faster for simplicial complex 3, this allows the coupling weight to be increased for the lower-dimensional interactions. We find this to be  somewhat counter intuitive---that is, as convergence through higher-dimensional interactions becomes faster, the optimal system relies more strongly on the lower-dimensional interactions.

\begin{figure}[t]
\includegraphics[width=\linewidth]{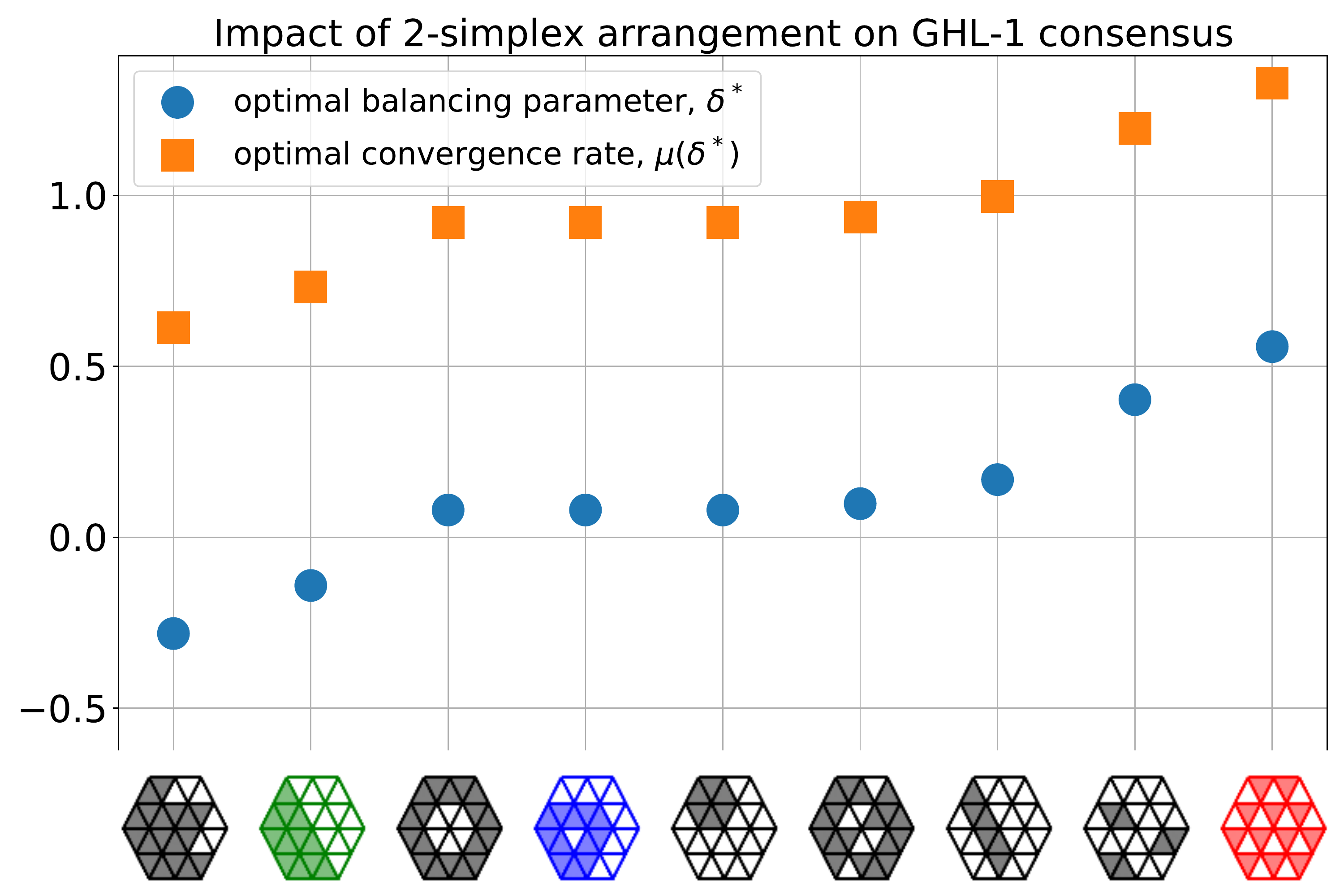}
\caption{
{\bf  2-simplex placement changes optimal convergence rate.} 
Visualization of the optimal convergence rate $\mu(\delta^*)$ and the optimal balancing parameter $\delta^*$ for different simplicial complexes that have the same nodes and edges but different locations and numbers of 2-simplices.
Interestingly, we do not identify a clear relation between the number of 2-simplices and $\mu(\delta^*)$ or $\delta^*$. However, we again find that convergence tends to be faster when the 2-simplices are more dispersed.
}
\label{fig:triangles}
\end{figure}

In Fig.~\ref{fig:triangles}, we further study the the optimal convergence rate $\mu(\delta^*)$ and optimal coupling $\delta^*$ for several additional simplicial complexes that have the same nodes and edges but different arrangements of 2-simplices.
We also now allow them to contain different numbers of 2-simplices, but we do not identify an obvious relation between that number and $\mu(\delta^*)$ or $\delta^*$.
That said, similar to Fig.~\ref{fig:exampleSC} we observe that faster convergence is associated with larger $\delta^*$.
However, it is important to stress that in general, this need not be the case. 
This phenomenon occurs here because we restricted our attention to
simplicial complexes that have the same nodes and edges, and thus the same gradient subspaces.

\section{Discussion}\label{sec:discussion}

We have presented here a model for consensus dynamics over simplicial complexes that uses a generalization of the Hodge Laplacian that we call \emph{generalized Hodge Laplacians}. It is likely that in many real-world applications of simplicial complex dynamics, higher- and lower-dimensional interactions between simplices are not equally weighted.  In these applications, the generalized Hodge Laplacian could prove a useful tool in modeling this phenomenon.

Our work identified a decoupling of consensus dynamics given by Eq.~\eqref{dynamics} onto the Hodge subspaces, and we  showed that the collective dynamics converge to the harmonic space (which is isomorphic to the homology group of the simplicial complex). 
These findings open several new direction of research in working with dynamics, consensus, and homology of simplicial complexes.  The literature on consensus and other dynamical processes over networks (e.g., diffusion) has been restricted to studying convergence toward a 1D subspace. Understanding how systems converge toward higher-dimensional subspaces represents an important new direction that may be foundational to more complicated dynamical systems. In particular, biological neurosystems are also likely to organize along low-dimensional subspaces and/or manifolds \cite{mcintosh2019hidden}, and it has been noted that deep neural networks are non-convergent and may even approach  limit cycles \cite{bock2019non} that are within subspaces of dimension greater than 1.

On the practical side, we also developed  theory to optimally balance higher- and lower-order interactions to maximally accelerate convergence. 
Interestingly, the optimum coincides with a balancing of dynamics on the gradient and curl subspaces of the Hodge decomposition. We additionally explored the effects of topological structure, finding that consensus over edges is accelerated when 2-simplices are well dispersed.
There is much to learn about the specific quantitative reasons for this behavior. Research may also be done into the optimal balancing parameter and convergence rate for randomly generated simplicial complexes, or those that satisfy certain criteria specific to neuroscience, machine learning, or social network applications.

Other possible future directions can be found by revisiting our modeling choices. While GHL consensus is defined for higher dimensions, we restricted our simulations to the $k=1$ case for simplicity.  It is worth investigating not only GHL-$k$ consensus for $k>1$, but also properties of the gradient, curl, and harmonic subspaces for these higher dimensions. 
At the same time, we introduced a balancing based on the combinatorial Hodge Laplacian, and it would also be beneficial to balance normalized Hodge Laplacians \cite{schaub2020random} (which were recently introduced  to model random walks on simplicial complexes). Our generalized Hodge Laplacian may be useful for other dynamics as well, and we are currently investigating the relation between our work and explosive synchronization on simplicial complexes \cite{millan2020explosive}.
Another interesting direction would be to consider adaptive dynamics in which the balancing parameter is allowed to change with time, i.e., $\delta(t)$.
Alternatively, 
we  could   introduce a learning rate $\beta(t)$ 
and replace Eq.~\eqref{dynamics} by
$\frac{d}{dt}{\bf x}(t) = -\beta(t) L^\delta_k {\bf x}(t) $. We could   
allow the learning rate to change over time, which is commonly done in decentralized machine learning.

Finally, we conclude by highlighting how our proposed model and theory may be useful to help support the development of decentralized algorithms for machine learning and artificial intelligence (AI). In this application, data is dispersed across a network of computer nodes (e.g., separate CPUs that are either organized into high-performance computer cluster or are spatially dispersed through a network of remote sensors). Such algorithms already rely on various forms of consensus in which the states of nodes (i.e., 1-simplices) coordinate to learn a single model for data. In this context, most research  focuses on designing algorithms that converge as fast and efficient as possible (see review \cite{nokleby2020scaling}). 
Our findings here suggest that it may be feasible to generalize such algorithms to   also coordinate the states of 2-simplices and higher-dimensional simplices. 
This potential new direction for decentralized learning is especially relevant, given this community's growing interest \cite{roddenberry2019hodgenet,bunch2020simplicial} to utilize simplicial complexes and Hodge Laplacians to design  deep neural networks.

Python code that implements GHL-1 consensus on simplicial complexes and reproduces our experiments can be found at \cite{cameron_code}.

\begin{acknowledgments}
CZ, HD, and DT were supported in part by the National Science Foundation (EDT-1551069). DT also acknowledges NSF grant DMS-2052720 and the Simons Foundation (grant \#578333).
PSS acknowledges NSF grant MCB-2126177.
\end{acknowledgments}

\bibliography{main}

\begin{thebibliography}{42}%
\makeatletter
\providecommand \@ifxundefined [1]{%
 \@ifx{#1\undefined}
}%
\providecommand \@ifnum [1]{%
 \ifnum #1\expandafter \@firstoftwo
 \else \expandafter \@secondoftwo
 \fi
}%
\providecommand \@ifx [1]{%
 \ifx #1\expandafter \@firstoftwo
 \else \expandafter \@secondoftwo
 \fi
}%
\providecommand \natexlab [1]{#1}%
\providecommand \enquote  [1]{``#1''}%
\providecommand \bibnamefont  [1]{#1}%
\providecommand \bibfnamefont [1]{#1}%
\providecommand \citenamefont [1]{#1}%
\providecommand \href@noop [0]{\@secondoftwo}%
\providecommand \href [0]{\begingroup \@sanitize@url \@href}%
\providecommand \@href[1]{\@@startlink{#1}\@@href}%
\providecommand \@@href[1]{\endgroup#1\@@endlink}%
\providecommand \@sanitize@url [0]{\catcode `\\12\catcode `\$12\catcode
  `\&12\catcode `\#12\catcode `\^12\catcode `\_12\catcode `\%12\relax}%
\providecommand \@@startlink[1]{}%
\providecommand \@@endlink[0]{}%
\providecommand \url  [0]{\begingroup\@sanitize@url \@url }%
\providecommand \@url [1]{\endgroup\@href {#1}{\urlprefix }}%
\providecommand \urlprefix  [0]{URL }%
\providecommand \Eprint [0]{\href }%
\providecommand \doibase [0]{http://dx.doi.org/}%
\providecommand \selectlanguage [0]{\@gobble}%
\providecommand \bibinfo  [0]{\@secondoftwo}%
\providecommand \bibfield  [0]{\@secondoftwo}%
\providecommand \translation [1]{[#1]}%
\providecommand \BibitemOpen [0]{}%
\providecommand \bibitemStop [0]{}%
\providecommand \bibitemNoStop [0]{.\EOS\space}%
\providecommand \EOS [0]{\spacefactor3000\relax}%
\providecommand \BibitemShut  [1]{\csname bibitem#1\endcsname}%
\let\auto@bib@innerbib\@empty
\bibitem [{\citenamefont {Schaub}\ \emph {et~al.}(2020)\citenamefont {Schaub},
  \citenamefont {Benson}, \citenamefont {Horn}, \citenamefont {Lippner},\ and\
  \citenamefont {Jadbabaie}}]{schaub2020random}%
  \BibitemOpen
  \bibfield  {author} {\bibinfo {author} {\bibfnamefont {M.~T.}\ \bibnamefont
  {Schaub}}, \bibinfo {author} {\bibfnamefont {A.~R.}\ \bibnamefont {Benson}},
  \bibinfo {author} {\bibfnamefont {P.}~\bibnamefont {Horn}}, \bibinfo {author}
  {\bibfnamefont {G.}~\bibnamefont {Lippner}}, \ and\ \bibinfo {author}
  {\bibfnamefont {A.}~\bibnamefont {Jadbabaie}},\ }\bibfield  {title} {\enquote
  {\bibinfo {title} {Random walks on simplicial complexes and the normalized
  hodge 1-laplacian},}\ }\href@noop {} {\bibfield  {journal} {\bibinfo
  {journal} {SIAM Review}\ }\textbf {\bibinfo {volume} {62}},\ \bibinfo {pages}
  {353--391} (\bibinfo {year} {2020})}\BibitemShut {NoStop}%
\bibitem [{\citenamefont {Mukherjee}\ and\ \citenamefont
  {Steenbergen}(2016)}]{mukherjee2016random}%
  \BibitemOpen
  \bibfield  {author} {\bibinfo {author} {\bibfnamefont {S.}~\bibnamefont
  {Mukherjee}}\ and\ \bibinfo {author} {\bibfnamefont {J.}~\bibnamefont
  {Steenbergen}},\ }\bibfield  {title} {\enquote {\bibinfo {title} {Random
  walks on simplicial complexes and harmonics},}\ }\href@noop {} {\bibfield
  {journal} {\bibinfo  {journal} {Random structures \& algorithms}\ }\textbf
  {\bibinfo {volume} {49}},\ \bibinfo {pages} {379--405} (\bibinfo {year}
  {2016})}\BibitemShut {NoStop}%
\bibitem [{\citenamefont {Rosenthal}(2014)}]{rosenthal2014simplicial}%
  \BibitemOpen
  \bibfield  {author} {\bibinfo {author} {\bibfnamefont {R.}~\bibnamefont
  {Rosenthal}},\ }\bibfield  {title} {\enquote {\bibinfo {title} {Simplicial
  branching random walks and their applications},}\ }\href@noop {} {\bibfield
  {journal} {\bibinfo  {journal} {arXiv preprint arXiv:1412.5406}\ } (\bibinfo
  {year} {2014})}\BibitemShut {NoStop}%
\bibitem [{\citenamefont {Iacopini}\ \emph {et~al.}(2019)\citenamefont
  {Iacopini}, \citenamefont {Petri}, \citenamefont {Barrat},\ and\
  \citenamefont {Latora}}]{iacopini2019simplicial}%
  \BibitemOpen
  \bibfield  {author} {\bibinfo {author} {\bibfnamefont {I.}~\bibnamefont
  {Iacopini}}, \bibinfo {author} {\bibfnamefont {G.}~\bibnamefont {Petri}},
  \bibinfo {author} {\bibfnamefont {A.}~\bibnamefont {Barrat}}, \ and\ \bibinfo
  {author} {\bibfnamefont {V.}~\bibnamefont {Latora}},\ }\bibfield  {title}
  {\enquote {\bibinfo {title} {Simplicial models of social contagion},}\
  }\href@noop {} {\bibfield  {journal} {\bibinfo  {journal} {Nature
  communications}\ }\textbf {\bibinfo {volume} {10}},\ \bibinfo {pages} {1--9}
  (\bibinfo {year} {2019})}\BibitemShut {NoStop}%
\bibitem [{\citenamefont {Matamalas}, \citenamefont {G{\'o}mez},\ and\
  \citenamefont {Arenas}(2020)}]{matamalas2020abrupt}%
  \BibitemOpen
  \bibfield  {author} {\bibinfo {author} {\bibfnamefont {J.~T.}\ \bibnamefont
  {Matamalas}}, \bibinfo {author} {\bibfnamefont {S.}~\bibnamefont
  {G{\'o}mez}}, \ and\ \bibinfo {author} {\bibfnamefont {A.}~\bibnamefont
  {Arenas}},\ }\bibfield  {title} {\enquote {\bibinfo {title} {Abrupt phase
  transition of epidemic spreading in simplicial complexes},}\ }\href@noop {}
  {\bibfield  {journal} {\bibinfo  {journal} {Physical Review Research}\
  }\textbf {\bibinfo {volume} {2}},\ \bibinfo {pages} {012049} (\bibinfo {year}
  {2020})}\BibitemShut {NoStop}%
\bibitem [{\citenamefont {Petri}\ \emph {et~al.}(2014)\citenamefont {Petri},
  \citenamefont {Expert}, \citenamefont {Turkheimer}, \citenamefont
  {Carhart-Harris}, \citenamefont {Nutt}, \citenamefont {Hellyer},\ and\
  \citenamefont {Vaccarino}}]{petri2014homological}%
  \BibitemOpen
  \bibfield  {author} {\bibinfo {author} {\bibfnamefont {G.}~\bibnamefont
  {Petri}}, \bibinfo {author} {\bibfnamefont {P.}~\bibnamefont {Expert}},
  \bibinfo {author} {\bibfnamefont {F.}~\bibnamefont {Turkheimer}}, \bibinfo
  {author} {\bibfnamefont {R.}~\bibnamefont {Carhart-Harris}}, \bibinfo
  {author} {\bibfnamefont {D.}~\bibnamefont {Nutt}}, \bibinfo {author}
  {\bibfnamefont {P.~J.}\ \bibnamefont {Hellyer}}, \ and\ \bibinfo {author}
  {\bibfnamefont {F.}~\bibnamefont {Vaccarino}},\ }\bibfield  {title} {\enquote
  {\bibinfo {title} {Homological scaffolds of brain functional networks},}\
  }\href@noop {} {\bibfield  {journal} {\bibinfo  {journal} {Journal of The
  Royal Society Interface}\ }\textbf {\bibinfo {volume} {11}},\ \bibinfo
  {pages} {20140873} (\bibinfo {year} {2014})}\BibitemShut {NoStop}%
\bibitem [{\citenamefont {Vergne}, \citenamefont {Decreusefond},\ and\
  \citenamefont {Martins}(2014)}]{vergne2014simplicial}%
  \BibitemOpen
  \bibfield  {author} {\bibinfo {author} {\bibfnamefont {A.}~\bibnamefont
  {Vergne}}, \bibinfo {author} {\bibfnamefont {L.}~\bibnamefont
  {Decreusefond}}, \ and\ \bibinfo {author} {\bibfnamefont {P.}~\bibnamefont
  {Martins}},\ }\bibfield  {title} {\enquote {\bibinfo {title} {Simplicial
  homology for future cellular networks},}\ }\href@noop {} {\bibfield
  {journal} {\bibinfo  {journal} {IEEE Transactions on Mobile Computing}\
  }\textbf {\bibinfo {volume} {14}},\ \bibinfo {pages} {1712--1725} (\bibinfo
  {year} {2014})}\BibitemShut {NoStop}%
\bibitem [{\citenamefont {Gambuzza}\ \emph {et~al.}(2020)\citenamefont
  {Gambuzza}, \citenamefont {Di~Patti}, \citenamefont {Gallo}, \citenamefont
  {Lepri}, \citenamefont {Romance}, \citenamefont {Criado}, \citenamefont
  {Frasca}, \citenamefont {Latora},\ and\ \citenamefont
  {Boccaletti}}]{gambuzza2020master}%
  \BibitemOpen
  \bibfield  {author} {\bibinfo {author} {\bibfnamefont {L.}~\bibnamefont
  {Gambuzza}}, \bibinfo {author} {\bibfnamefont {F.}~\bibnamefont {Di~Patti}},
  \bibinfo {author} {\bibfnamefont {L.}~\bibnamefont {Gallo}}, \bibinfo
  {author} {\bibfnamefont {S.}~\bibnamefont {Lepri}}, \bibinfo {author}
  {\bibfnamefont {M.}~\bibnamefont {Romance}}, \bibinfo {author} {\bibfnamefont
  {R.}~\bibnamefont {Criado}}, \bibinfo {author} {\bibfnamefont
  {M.}~\bibnamefont {Frasca}}, \bibinfo {author} {\bibfnamefont
  {V.}~\bibnamefont {Latora}}, \ and\ \bibinfo {author} {\bibfnamefont
  {S.}~\bibnamefont {Boccaletti}},\ }\bibfield  {title} {\enquote {\bibinfo
  {title} {The master stability function for synchronization in simplicial
  complexes},}\ }\href@noop {} {\bibfield  {journal} {\bibinfo  {journal}
  {arXiv preprint arXiv:2004.03913}\ } (\bibinfo {year} {2020})}\BibitemShut
  {NoStop}%
\bibitem [{\citenamefont {Skardal}\ and\ \citenamefont
  {Arenas}(2019)}]{skardal2019abrupt}%
  \BibitemOpen
  \bibfield  {author} {\bibinfo {author} {\bibfnamefont {P.~S.}\ \bibnamefont
  {Skardal}}\ and\ \bibinfo {author} {\bibfnamefont {A.}~\bibnamefont
  {Arenas}},\ }\bibfield  {title} {\enquote {\bibinfo {title} {Abrupt
  desynchronization and extensive multistability in globally coupled oscillator
  simplexes},}\ }\href@noop {} {\bibfield  {journal} {\bibinfo  {journal}
  {Physical Review Letters}\ }\textbf {\bibinfo {volume} {122}},\ \bibinfo
  {pages} {248301} (\bibinfo {year} {2019})}\BibitemShut {NoStop}%
\bibitem [{\citenamefont {Arnaudon}\ \emph {et~al.}(2021)\citenamefont
  {Arnaudon}, \citenamefont {Peach}, \citenamefont {Petri},\ and\ \citenamefont
  {Expert}}]{arnaudon2021connecting}%
  \BibitemOpen
  \bibfield  {author} {\bibinfo {author} {\bibfnamefont {A.}~\bibnamefont
  {Arnaudon}}, \bibinfo {author} {\bibfnamefont {R.~L.}\ \bibnamefont {Peach}},
  \bibinfo {author} {\bibfnamefont {G.}~\bibnamefont {Petri}}, \ and\ \bibinfo
  {author} {\bibfnamefont {P.}~\bibnamefont {Expert}},\ }\href@noop {}
  {\enquote {\bibinfo {title} {Connecting hodge and sakaguchi-kuramoto: a
  mathematical framework for coupled oscillators on simplicial complexes},}\ }
  (\bibinfo {year} {2021}),\ \Eprint {http://arxiv.org/abs/2111.11073}
  {arXiv:2111.11073 [math-ph]} \BibitemShut {NoStop}%
\bibitem [{\citenamefont {Maleti{\'c}}\ and\ \citenamefont
  {Rajkovi{\'c}}(2014)}]{maletic2014consensus}%
  \BibitemOpen
  \bibfield  {author} {\bibinfo {author} {\bibfnamefont {S.}~\bibnamefont
  {Maleti{\'c}}}\ and\ \bibinfo {author} {\bibfnamefont {M.}~\bibnamefont
  {Rajkovi{\'c}}},\ }\bibfield  {title} {\enquote {\bibinfo {title} {Consensus
  formation on a simplicial complex of opinions},}\ }\href@noop {} {\bibfield
  {journal} {\bibinfo  {journal} {Physica A: Statistical Mechanics and its
  Applications}\ }\textbf {\bibinfo {volume} {397}},\ \bibinfo {pages}
  {111--120} (\bibinfo {year} {2014})}\BibitemShut {NoStop}%
\bibitem [{\citenamefont {Neuh{\"a}user}, \citenamefont {Mellor},\ and\
  \citenamefont {Lambiotte}(2020)}]{neuhauser2020multibody}%
  \BibitemOpen
  \bibfield  {author} {\bibinfo {author} {\bibfnamefont {L.}~\bibnamefont
  {Neuh{\"a}user}}, \bibinfo {author} {\bibfnamefont {A.}~\bibnamefont
  {Mellor}}, \ and\ \bibinfo {author} {\bibfnamefont {R.}~\bibnamefont
  {Lambiotte}},\ }\bibfield  {title} {\enquote {\bibinfo {title} {Multibody
  interactions and nonlinear consensus dynamics on networked systems},}\
  }\href@noop {} {\bibfield  {journal} {\bibinfo  {journal} {Physical Review
  E}\ }\textbf {\bibinfo {volume} {101}},\ \bibinfo {pages} {032310} (\bibinfo
  {year} {2020})}\BibitemShut {NoStop}%
\bibitem [{\citenamefont {Arai}, \citenamefont {Brandt},\ and\ \citenamefont
  {Dabaghian}(2014)}]{arai2014effects}%
  \BibitemOpen
  \bibfield  {author} {\bibinfo {author} {\bibfnamefont {M.}~\bibnamefont
  {Arai}}, \bibinfo {author} {\bibfnamefont {V.}~\bibnamefont {Brandt}}, \ and\
  \bibinfo {author} {\bibfnamefont {Y.}~\bibnamefont {Dabaghian}},\ }\bibfield
  {title} {\enquote {\bibinfo {title} {The effects of theta precession on
  spatial learning and simplicial complex dynamics in a topological model of
  the hippocampal spatial map},}\ }\href@noop {} {\bibfield  {journal}
  {\bibinfo  {journal} {PLoS computational biology}\ }\textbf {\bibinfo
  {volume} {10}},\ \bibinfo {pages} {e1003651} (\bibinfo {year}
  {2014})}\BibitemShut {NoStop}%
\bibitem [{\citenamefont {Roddenberry}\ and\ \citenamefont
  {Segarra}(2019)}]{roddenberry2019hodgenet}%
  \BibitemOpen
  \bibfield  {author} {\bibinfo {author} {\bibfnamefont {T.~M.}\ \bibnamefont
  {Roddenberry}}\ and\ \bibinfo {author} {\bibfnamefont {S.}~\bibnamefont
  {Segarra}},\ }\bibfield  {title} {\enquote {\bibinfo {title} {Hodgenet: Graph
  neural networks for edge data},}\ }in\ \href@noop {} {\emph {\bibinfo
  {booktitle} {2019 53rd Asilomar Conference on Signals, Systems, and
  Computers}}}\ (\bibinfo {organization} {IEEE},\ \bibinfo {year} {2019})\ pp.\
  \bibinfo {pages} {220--224}\BibitemShut {NoStop}%
\bibitem [{\citenamefont {Bunch}\ \emph {et~al.}(2020)\citenamefont {Bunch},
  \citenamefont {You}, \citenamefont {Fung},\ and\ \citenamefont
  {Singh}}]{bunch2020simplicial}%
  \BibitemOpen
  \bibfield  {author} {\bibinfo {author} {\bibfnamefont {E.}~\bibnamefont
  {Bunch}}, \bibinfo {author} {\bibfnamefont {Q.}~\bibnamefont {You}}, \bibinfo
  {author} {\bibfnamefont {G.}~\bibnamefont {Fung}}, \ and\ \bibinfo {author}
  {\bibfnamefont {V.}~\bibnamefont {Singh}},\ }\bibfield  {title} {\enquote
  {\bibinfo {title} {Simplicial 2-complex convolutional neural nets},}\
  }\href@noop {} {\bibfield  {journal} {\bibinfo  {journal} {arXiv preprint
  arXiv:2012.06010}\ } (\bibinfo {year} {2020})}\BibitemShut {NoStop}%
\bibitem [{\citenamefont {Schaub}\ \emph {et~al.}(2021)\citenamefont {Schaub},
  \citenamefont {Seby}, \citenamefont {Frantzen}, \citenamefont {Roddenberry},
  \citenamefont {Zhu},\ and\ \citenamefont {Segarra}}]{schaub2021signal}%
  \BibitemOpen
  \bibfield  {author} {\bibinfo {author} {\bibfnamefont {M.~T.}\ \bibnamefont
  {Schaub}}, \bibinfo {author} {\bibfnamefont {J.-B.}\ \bibnamefont {Seby}},
  \bibinfo {author} {\bibfnamefont {F.}~\bibnamefont {Frantzen}}, \bibinfo
  {author} {\bibfnamefont {T.~M.}\ \bibnamefont {Roddenberry}}, \bibinfo
  {author} {\bibfnamefont {Y.}~\bibnamefont {Zhu}}, \ and\ \bibinfo {author}
  {\bibfnamefont {S.}~\bibnamefont {Segarra}},\ }\bibfield  {title} {\enquote
  {\bibinfo {title} {Signal processing on simplicial complexes},}\ }\href@noop
  {} {\bibfield  {journal} {\bibinfo  {journal} {arXiv preprint
  arXiv:2106.07471}\ } (\bibinfo {year} {2021})}\BibitemShut {NoStop}%
\bibitem [{\citenamefont {Gleeson}\ \emph {et~al.}(2012)\citenamefont
  {Gleeson}, \citenamefont {Melnik}, \citenamefont {Ward}, \citenamefont
  {Porter},\ and\ \citenamefont {Mucha}}]{gleeson2012accuracy}%
  \BibitemOpen
  \bibfield  {author} {\bibinfo {author} {\bibfnamefont {J.~P.}\ \bibnamefont
  {Gleeson}}, \bibinfo {author} {\bibfnamefont {S.}~\bibnamefont {Melnik}},
  \bibinfo {author} {\bibfnamefont {J.~A.}\ \bibnamefont {Ward}}, \bibinfo
  {author} {\bibfnamefont {M.~A.}\ \bibnamefont {Porter}}, \ and\ \bibinfo
  {author} {\bibfnamefont {P.~J.}\ \bibnamefont {Mucha}},\ }\bibfield  {title}
  {\enquote {\bibinfo {title} {Accuracy of mean-field theory for dynamics on
  real-world networks},}\ }\href@noop {} {\bibfield  {journal} {\bibinfo
  {journal} {Physical Review E}\ }\textbf {\bibinfo {volume} {85}},\ \bibinfo
  {pages} {026106} (\bibinfo {year} {2012})}\BibitemShut {NoStop}%
\bibitem [{\citenamefont {Snijders}, \citenamefont {Van~de Bunt},\ and\
  \citenamefont {Steglich}(2010)}]{snijders2010introduction}%
  \BibitemOpen
  \bibfield  {author} {\bibinfo {author} {\bibfnamefont {T.~A.}\ \bibnamefont
  {Snijders}}, \bibinfo {author} {\bibfnamefont {G.~G.}\ \bibnamefont {Van~de
  Bunt}}, \ and\ \bibinfo {author} {\bibfnamefont {C.~E.}\ \bibnamefont
  {Steglich}},\ }\bibfield  {title} {\enquote {\bibinfo {title} {Introduction
  to stochastic actor-based models for network dynamics},}\ }\href@noop {}
  {\bibfield  {journal} {\bibinfo  {journal} {Social networks}\ }\textbf
  {\bibinfo {volume} {32}},\ \bibinfo {pages} {44--60} (\bibinfo {year}
  {2010})}\BibitemShut {NoStop}%
\bibitem [{\citenamefont {Bardin}, \citenamefont {Spreemann},\ and\
  \citenamefont {Hess}(2019)}]{bardin2019topological}%
  \BibitemOpen
  \bibfield  {author} {\bibinfo {author} {\bibfnamefont {J.-B.}\ \bibnamefont
  {Bardin}}, \bibinfo {author} {\bibfnamefont {G.}~\bibnamefont {Spreemann}}, \
  and\ \bibinfo {author} {\bibfnamefont {K.}~\bibnamefont {Hess}},\ }\bibfield
  {title} {\enquote {\bibinfo {title} {Topological exploration of artificial
  neuronal network dynamics},}\ }\href@noop {} {\bibfield  {journal} {\bibinfo
  {journal} {Network Neuroscience}\ }\textbf {\bibinfo {volume} {3}},\ \bibinfo
  {pages} {725--743} (\bibinfo {year} {2019})}\BibitemShut {NoStop}%
\bibitem [{\citenamefont {Hatcher}(2001)}]{hatcher2005algebraic}%
  \BibitemOpen
  \bibfield  {author} {\bibinfo {author} {\bibfnamefont {A.}~\bibnamefont
  {Hatcher}},\ }\href@noop {} {\emph {\bibinfo {title} {Algebraic topology}}}\
  (\bibinfo  {publisher} {Cambridge University Press},\ \bibinfo {year}
  {2001})\BibitemShut {NoStop}%
\bibitem [{\citenamefont {Muhammad}\ and\ \citenamefont
  {Egerstedt}(2006)}]{muhammad2006control}%
  \BibitemOpen
  \bibfield  {author} {\bibinfo {author} {\bibfnamefont {A.}~\bibnamefont
  {Muhammad}}\ and\ \bibinfo {author} {\bibfnamefont {M.}~\bibnamefont
  {Egerstedt}},\ }\bibfield  {title} {\enquote {\bibinfo {title} {Control using
  higher order laplacians in network topologies},}\ }in\ \href@noop {} {\emph
  {\bibinfo {booktitle} {Proc. of 17th International Symposium on Mathematical
  Theory of Networks and Systems}}}\ (\bibinfo {organization} {Citeseer},\
  \bibinfo {year} {2006})\ pp.\ \bibinfo {pages} {1024--1038}\BibitemShut
  {NoStop}%
\bibitem [{\citenamefont {Hinsz}(1990)}]{hinsz1990cognitive}%
  \BibitemOpen
  \bibfield  {author} {\bibinfo {author} {\bibfnamefont {V.~B.}\ \bibnamefont
  {Hinsz}},\ }\bibfield  {title} {\enquote {\bibinfo {title} {Cognitive and
  consensus processes in group recognition memory performance.}}\ }\href@noop
  {} {\bibfield  {journal} {\bibinfo  {journal} {Journal of Personality and
  Social Psychology}\ }\textbf {\bibinfo {volume} {59}},\ \bibinfo {pages}
  {705} (\bibinfo {year} {1990})}\BibitemShut {NoStop}%
\bibitem [{\citenamefont {Fiol}(1994)}]{fiol1994consensus}%
  \BibitemOpen
  \bibfield  {author} {\bibinfo {author} {\bibfnamefont {C.~M.}\ \bibnamefont
  {Fiol}},\ }\bibfield  {title} {\enquote {\bibinfo {title} {Consensus,
  diversity, and learning in organizations},}\ }\href@noop {} {\bibfield
  {journal} {\bibinfo  {journal} {Organization Science}\ }\textbf {\bibinfo
  {volume} {5}},\ \bibinfo {pages} {403--420} (\bibinfo {year}
  {1994})}\BibitemShut {NoStop}%
\bibitem [{\citenamefont {Conradt}\ and\ \citenamefont
  {Roper}(2005)}]{conradt2005consensus}%
  \BibitemOpen
  \bibfield  {author} {\bibinfo {author} {\bibfnamefont {L.}~\bibnamefont
  {Conradt}}\ and\ \bibinfo {author} {\bibfnamefont {T.~J.}\ \bibnamefont
  {Roper}},\ }\bibfield  {title} {\enquote {\bibinfo {title} {Consensus
  decision making in animals},}\ }\href@noop {} {\bibfield  {journal} {\bibinfo
   {journal} {Trends in Ecology \& Evolution}\ }\textbf {\bibinfo {volume}
  {20}},\ \bibinfo {pages} {449--456} (\bibinfo {year} {2005})}\BibitemShut
  {NoStop}%
\bibitem [{\citenamefont {Song}\ and\ \citenamefont
  {Taylor}(2021)}]{song2021asymmetric}%
  \BibitemOpen
  \bibfield  {author} {\bibinfo {author} {\bibfnamefont {Z.}~\bibnamefont
  {Song}}\ and\ \bibinfo {author} {\bibfnamefont {D.}~\bibnamefont {Taylor}},\
  }\bibfield  {title} {\enquote {\bibinfo {title} {Asymmetric coupling
  optimizes interconnected consensus systems},}\ }\href@noop {} {\bibfield
  {journal} {\bibinfo  {journal} {arXiv preprint arXiv:2106.13127}\ } (\bibinfo
  {year} {2021})}\BibitemShut {NoStop}%
\bibitem [{\citenamefont {Tsitsiklis}, \citenamefont {Bertsekas},\ and\
  \citenamefont {Athans}(1986)}]{tsitsiklis1986distributed}%
  \BibitemOpen
  \bibfield  {author} {\bibinfo {author} {\bibfnamefont {J.}~\bibnamefont
  {Tsitsiklis}}, \bibinfo {author} {\bibfnamefont {D.}~\bibnamefont
  {Bertsekas}}, \ and\ \bibinfo {author} {\bibfnamefont {M.}~\bibnamefont
  {Athans}},\ }\bibfield  {title} {\enquote {\bibinfo {title} {Distributed
  asynchronous deterministic and stochastic gradient optimization
  algorithms},}\ }\href@noop {} {\bibfield  {journal} {\bibinfo  {journal}
  {IEEE transactions on automatic control}\ }\textbf {\bibinfo {volume} {31}},\
  \bibinfo {pages} {803--812} (\bibinfo {year} {1986})}\BibitemShut {NoStop}%
\bibitem [{\citenamefont {Boyd}\ \emph {et~al.}(2005)\citenamefont {Boyd},
  \citenamefont {Ghosh}, \citenamefont {Prabhakar},\ and\ \citenamefont
  {Shah}}]{boyd2005gossip}%
  \BibitemOpen
  \bibfield  {author} {\bibinfo {author} {\bibfnamefont {S.}~\bibnamefont
  {Boyd}}, \bibinfo {author} {\bibfnamefont {A.}~\bibnamefont {Ghosh}},
  \bibinfo {author} {\bibfnamefont {B.}~\bibnamefont {Prabhakar}}, \ and\
  \bibinfo {author} {\bibfnamefont {D.}~\bibnamefont {Shah}},\ }\bibfield
  {title} {\enquote {\bibinfo {title} {Gossip algorithms: Design, analysis and
  applications},}\ }in\ \href@noop {} {\emph {\bibinfo {booktitle} {Proceedings
  IEEE 24th Annual Joint Conference of the IEEE Computer and Communications
  Societies.}}},\ Vol.~\bibinfo {volume} {3}\ (\bibinfo {organization} {IEEE},\
  \bibinfo {year} {2005})\ pp.\ \bibinfo {pages} {1653--1664}\BibitemShut
  {NoStop}%
\bibitem [{\citenamefont {Bijral}, \citenamefont {Sarwate},\ and\ \citenamefont
  {Srebro}(2017)}]{bijral2017data}%
  \BibitemOpen
  \bibfield  {author} {\bibinfo {author} {\bibfnamefont {A.~S.}\ \bibnamefont
  {Bijral}}, \bibinfo {author} {\bibfnamefont {A.~D.}\ \bibnamefont {Sarwate}},
  \ and\ \bibinfo {author} {\bibfnamefont {N.}~\bibnamefont {Srebro}},\
  }\bibfield  {title} {\enquote {\bibinfo {title} {Data-dependent convergence
  for consensus stochastic optimization},}\ }\href@noop {} {\bibfield
  {journal} {\bibinfo  {journal} {IEEE Transactions on Automatic Control}\
  }\textbf {\bibinfo {volume} {62}},\ \bibinfo {pages} {4483--4498} (\bibinfo
  {year} {2017})}\BibitemShut {NoStop}%
\bibitem [{\citenamefont {Huynh}, \citenamefont {Dutta},\ and\ \citenamefont
  {Taylor}(2021)}]{huynh2021}%
  \BibitemOpen
  \bibfield  {author} {\bibinfo {author} {\bibfnamefont {B.}~\bibnamefont
  {Huynh}}, \bibinfo {author} {\bibfnamefont {H.}~\bibnamefont {Dutta}}, \ and\
  \bibinfo {author} {\bibfnamefont {D.}~\bibnamefont {Taylor}},\ }\bibfield
  {title} {\enquote {\bibinfo {title} {Impact of community structure on
  consensus machine learning},}\ }\href@noop {} {\bibfield  {journal} {\bibinfo
   {journal} {arXiv preprint arXiv:2011.01334}\ } (\bibinfo {year}
  {2021})}\BibitemShut {NoStop}%
\bibitem [{\citenamefont {Assran}\ \emph {et~al.}(2019)\citenamefont {Assran},
  \citenamefont {Loizou}, \citenamefont {Ballas},\ and\ \citenamefont
  {Rabbat}}]{assran2019stochastic}%
  \BibitemOpen
  \bibfield  {author} {\bibinfo {author} {\bibfnamefont {M.}~\bibnamefont
  {Assran}}, \bibinfo {author} {\bibfnamefont {N.}~\bibnamefont {Loizou}},
  \bibinfo {author} {\bibfnamefont {N.}~\bibnamefont {Ballas}}, \ and\ \bibinfo
  {author} {\bibfnamefont {M.}~\bibnamefont {Rabbat}},\ }\bibfield  {title}
  {\enquote {\bibinfo {title} {Stochastic gradient push for distributed deep
  learning},}\ }in\ \href@noop {} {\emph {\bibinfo {booktitle} {International
  Conference on Machine Learning}}}\ (\bibinfo {organization} {PMLR},\ \bibinfo
  {year} {2019})\ pp.\ \bibinfo {pages} {344--353}\BibitemShut {NoStop}%
\bibitem [{\citenamefont {Niwa}\ \emph {et~al.}(2020)\citenamefont {Niwa},
  \citenamefont {Harada}, \citenamefont {Zhang},\ and\ \citenamefont
  {Kleijn}}]{niwa2020edge}%
  \BibitemOpen
  \bibfield  {author} {\bibinfo {author} {\bibfnamefont {K.}~\bibnamefont
  {Niwa}}, \bibinfo {author} {\bibfnamefont {N.}~\bibnamefont {Harada}},
  \bibinfo {author} {\bibfnamefont {G.}~\bibnamefont {Zhang}}, \ and\ \bibinfo
  {author} {\bibfnamefont {W.~B.}\ \bibnamefont {Kleijn}},\ }\bibfield  {title}
  {\enquote {\bibinfo {title} {Edge-consensus learning: Deep learning on p2p
  networks with nonhomogeneous data},}\ }in\ \href@noop {} {\emph {\bibinfo
  {booktitle} {Proceedings of the 26th ACM SIGKDD International Conference on
  Knowledge Discovery \& Data Mining}}}\ (\bibinfo {year} {2020})\ pp.\
  \bibinfo {pages} {668--678}\BibitemShut {NoStop}%
\bibitem [{\citenamefont {Vogels}, \citenamefont {Karimireddy},\ and\
  \citenamefont {Jaggi}(2020)}]{vogels2020powergossip}%
  \BibitemOpen
  \bibfield  {author} {\bibinfo {author} {\bibfnamefont {T.}~\bibnamefont
  {Vogels}}, \bibinfo {author} {\bibfnamefont {S.~P.}\ \bibnamefont
  {Karimireddy}}, \ and\ \bibinfo {author} {\bibfnamefont {M.}~\bibnamefont
  {Jaggi}},\ }\bibfield  {title} {\enquote {\bibinfo {title} {Powergossip:
  Practical low-rank communication compression in decentralized deep
  learning},}\ }\href@noop {} {\bibfield  {journal} {\bibinfo  {journal} {arXiv
  preprint arXiv:2008.01425}\ } (\bibinfo {year} {2020})}\BibitemShut {NoStop}%
\bibitem [{\citenamefont {Kong}\ \emph {et~al.}(2021)\citenamefont {Kong},
  \citenamefont {Lin}, \citenamefont {Koloskova}, \citenamefont {Jaggi},\ and\
  \citenamefont {Stich}}]{kong2021consensus}%
  \BibitemOpen
  \bibfield  {author} {\bibinfo {author} {\bibfnamefont {L.}~\bibnamefont
  {Kong}}, \bibinfo {author} {\bibfnamefont {T.}~\bibnamefont {Lin}}, \bibinfo
  {author} {\bibfnamefont {A.}~\bibnamefont {Koloskova}}, \bibinfo {author}
  {\bibfnamefont {M.}~\bibnamefont {Jaggi}}, \ and\ \bibinfo {author}
  {\bibfnamefont {S.~U.}\ \bibnamefont {Stich}},\ }\bibfield  {title} {\enquote
  {\bibinfo {title} {Consensus control for decentralized deep learning},}\
  }\href@noop {} {\bibfield  {journal} {\bibinfo  {journal} {arXiv preprint
  arXiv:2102.04828}\ } (\bibinfo {year} {2021})}\BibitemShut {NoStop}%
\bibitem [{\citenamefont {Bianconi}(2021)}]{bianconi2021topological}%
  \BibitemOpen
  \bibfield  {author} {\bibinfo {author} {\bibfnamefont {G.}~\bibnamefont
  {Bianconi}},\ }\bibfield  {title} {\enquote {\bibinfo {title} {The
  topological dirac equation of networks and simplicial complexes},}\
  }\href@noop {} {\bibfield  {journal} {\bibinfo  {journal} {arXiv preprint
  arXiv:2106.02929}\ } (\bibinfo {year} {2021})}\BibitemShut {NoStop}%
\bibitem [{\citenamefont {Mill{\'a}n}, \citenamefont {Torres},\ and\
  \citenamefont {Bianconi}(2020)}]{millan2020explosive}%
  \BibitemOpen
  \bibfield  {author} {\bibinfo {author} {\bibfnamefont {A.~P.}\ \bibnamefont
  {Mill{\'a}n}}, \bibinfo {author} {\bibfnamefont {J.~J.}\ \bibnamefont
  {Torres}}, \ and\ \bibinfo {author} {\bibfnamefont {G.}~\bibnamefont
  {Bianconi}},\ }\bibfield  {title} {\enquote {\bibinfo {title} {Explosive
  higher-order kuramoto dynamics on simplicial complexes},}\ }\href@noop {}
  {\bibfield  {journal} {\bibinfo  {journal} {Physical Review Letters}\
  }\textbf {\bibinfo {volume} {124}},\ \bibinfo {pages} {218301} (\bibinfo
  {year} {2020})}\BibitemShut {NoStop}%
\bibitem [{\citenamefont {Lim}(2015)}]{lim2015hodge}%
  \BibitemOpen
  \bibfield  {author} {\bibinfo {author} {\bibfnamefont {L.-H.}\ \bibnamefont
  {Lim}},\ }\bibfield  {title} {\enquote {\bibinfo {title} {Hodge laplacians on
  graphs},}\ }\href@noop {} {\bibfield  {journal} {\bibinfo  {journal} {arXiv
  preprint arXiv:1507.05379}\ } (\bibinfo {year} {2015})}\BibitemShut {NoStop}%
\bibitem [{\citenamefont {Lim}(2020)}]{lim2020hodge}%
  \BibitemOpen
  \bibfield  {author} {\bibinfo {author} {\bibfnamefont {L.-H.}\ \bibnamefont
  {Lim}},\ }\bibfield  {title} {\enquote {\bibinfo {title} {Hodge laplacians on
  graphs},}\ }\href@noop {} {\bibfield  {journal} {\bibinfo  {journal} {Siam
  Review}\ }\textbf {\bibinfo {volume} {62}},\ \bibinfo {pages} {685--715}
  (\bibinfo {year} {2020})}\BibitemShut {NoStop}%
\bibitem [{\citenamefont {Olfati-Saber}\ and\ \citenamefont
  {Murray}(2004)}]{olfati2004consensus}%
  \BibitemOpen
  \bibfield  {author} {\bibinfo {author} {\bibfnamefont {R.}~\bibnamefont
  {Olfati-Saber}}\ and\ \bibinfo {author} {\bibfnamefont {R.~M.}\ \bibnamefont
  {Murray}},\ }\bibfield  {title} {\enquote {\bibinfo {title} {Consensus
  problems in networks of agents with switching topology and time-delays},}\
  }\href@noop {} {\bibfield  {journal} {\bibinfo  {journal} {IEEE Transactions
  on automatic control}\ }\textbf {\bibinfo {volume} {49}},\ \bibinfo {pages}
  {1520--1533} (\bibinfo {year} {2004})}\BibitemShut {NoStop}%
\bibitem [{\citenamefont {McIntosh}\ and\ \citenamefont
  {Jirsa}(2019)}]{mcintosh2019hidden}%
  \BibitemOpen
  \bibfield  {author} {\bibinfo {author} {\bibfnamefont {A.~R.}\ \bibnamefont
  {McIntosh}}\ and\ \bibinfo {author} {\bibfnamefont {V.~K.}\ \bibnamefont
  {Jirsa}},\ }\bibfield  {title} {\enquote {\bibinfo {title} {The hidden
  repertoire of brain dynamics and dysfunction},}\ }\href@noop {} {\bibfield
  {journal} {\bibinfo  {journal} {Network Neuroscience}\ }\textbf {\bibinfo
  {volume} {3}},\ \bibinfo {pages} {994--1008} (\bibinfo {year}
  {2019})}\BibitemShut {NoStop}%
\bibitem [{\citenamefont {Bock}\ and\ \citenamefont
  {Wei{\ss}}(2019)}]{bock2019non}%
  \BibitemOpen
  \bibfield  {author} {\bibinfo {author} {\bibfnamefont {S.}~\bibnamefont
  {Bock}}\ and\ \bibinfo {author} {\bibfnamefont {M.}~\bibnamefont
  {Wei{\ss}}},\ }\bibfield  {title} {\enquote {\bibinfo {title}
  {Non-convergence and limit cycles in the adam optimizer},}\ }in\ \href@noop
  {} {\emph {\bibinfo {booktitle} {International Conference on Artificial
  Neural Networks}}}\ (\bibinfo {organization} {Springer},\ \bibinfo {year}
  {2019})\ pp.\ \bibinfo {pages} {232--243}\BibitemShut {NoStop}%
\bibitem [{\citenamefont {Nokleby}, \citenamefont {Raja},\ and\ \citenamefont
  {Bajwa}(2020)}]{nokleby2020scaling}%
  \BibitemOpen
  \bibfield  {author} {\bibinfo {author} {\bibfnamefont {M.}~\bibnamefont
  {Nokleby}}, \bibinfo {author} {\bibfnamefont {H.}~\bibnamefont {Raja}}, \
  and\ \bibinfo {author} {\bibfnamefont {W.~U.}\ \bibnamefont {Bajwa}},\
  }\bibfield  {title} {\enquote {\bibinfo {title} {Scaling-up distributed
  processing of data streams for machine learning},}\ }\href@noop {} {\bibfield
   {journal} {\bibinfo  {journal} {Proceedings of the IEEE}\ }\textbf {\bibinfo
  {volume} {108}},\ \bibinfo {pages} {1984--2012} (\bibinfo {year}
  {2020})}\BibitemShut {NoStop}%
\bibitem [{\citenamefont {Ziegler}(2021)}]{cameron_code}%
  \BibitemOpen
  \bibfield  {author} {\bibinfo {author} {\bibfnamefont {C.}~\bibnamefont
  {Ziegler}},\ }\href@noop {} {\enquote {\bibinfo {title} {Codebase for
  {B}alanced {H}odge {L}aplacians optimize consensus over simplicial complexes
  \url{https://github.com/cameronziegler/Balanced-Hodge-Laplacian}},}\ }
  (\bibinfo {year} {2021})\BibitemShut {NoStop}%
\end{thebibliography}%

\end{document}